\documentclass[12pt]{article}

\usepackage{jheppub}
\usepackage[utf8]{inputenc}

\usepackage[T1]{fontenc} 

\newcommand{\beq}{\begin{equation}}
\newcommand{\eeq}{\end{equation}}
\newcommand{\bal}{\begin{aligned}}
\newcommand{\eal}{\end{aligned}}
\newcommand{\rmd}{\mathrm d}

\usepackage{physics, xparse, tikz}
\usetikzlibrary{shapes.geometric}
\usetikzlibrary{decorations.pathmorphing}

\usepackage[normalem]{ulem}
\usepackage{cancel,xcolor}

\title{An Outsider's Perspective on \\Information Recovery in de Sitter Space}
\author{Lars Aalsma$^{a}$,}
\author{Sergio E. Aguilar-Gutierrez$^{b}$,}
\author{Watse Sybesma$^{c}$}
\emailAdd{laalsma@asu.edu}
\emailAdd{sergio.ernesto.aguilar@gmail.com}
\emailAdd{watse@hi.is}
\affiliation{$^a$Beyond: Center for Fundamental Concepts in Science, Arizona State University, Tempe, Arizona 85287, USA}
\affiliation{$^b$Institute for Theoretical Physics, KU Leuven,
Celestijnenlaan 200D, B-3001 Leuven, Belgium}
\affiliation{$^c$Science Institute, University of Iceland,
Dunhaga 3, 107 Reykjavík, Iceland}

\abstract{Entanglement islands play a crucial role in our understanding of how Hawking radiation encodes information in a black hole, but their relevance in cosmological spacetimes is less clear. In this paper, we continue our investigation of information recovery in de Sitter space and construct a two-dimensional model of gravity containing a domain wall that interpolates between de Sitter space and Rindler space. The Rindler wedges introduce weakly-gravitating asymptotic regions from which de Sitter space can be probed, yielding an outside perspective of the cosmological horizon. In contrast to earlier works, backreaction effects are under control by considering a quantum state that only breaks the thermal equilibrium of the Bunch-Davies state for a finite time. This allows information to be decoded from the Gibbons-Hawking radiation in a controlled fashion.}

\begin{document}

\maketitle

\section{Introduction}
Entanglement entropy has been a useful probe to learn about aspects of quantum gravity. Most famously, these developments have in recent years led to an expression for the fine-grained entropy of a subsystem that is entangled with a gravitating region \cite{Engelhardt:2014gca}. Arguably, one of the most interesting applications of these results has been to compute the entropy of Hawking radiation emitted by black holes \cite{Penington:2019npb,Almheiri:2019psf}. The results of these works are striking, as they reproduce a unitary Page curve \cite{Page:1993wv}, paving a way forward to understanding the physical mechanism that resolves the information paradox.\footnote{Strictly speaking, most explicit computations have been performed in models of two-dimensional gravity. Higher-dimensional generalizations of these results have been a subject of debate, see e.g.\cite{Almheiri:2019psy,Laddha:2020kvp,Geng:2020qvw,Chowdhury:2020hse,Geng:2020fxl,Geng:2021hlu,Chowdhury:2021nxw,Raju:2021lwh,Bousso:2022hlz}.} 

At the same time, these developments also have the potential to impact cosmology. In particular, a cosmological horizon can be associated a finite entropy, like a black hole. In a sense that we will make precise, this gives rise to a cosmological version of the information paradox. It is then natural to expect that entanglement islands could resolve this paradox and allow information to be decoded from the Gibbons-Hawking radiation. In this work, we are interested in formulating this question in de Sitter space from the perspective of the static patch. Contrasting to earlier work on this topic, such as \cite{Aalsma:2021bit,Kames-King:2021etp}, we argue that in the particular two-dimensional gravity model we study it is possible to decode information and at the same time avoid catastrophic backreaction.

Before we discuss the details, let us first mention some important facts about entropy in de Sitter space. The reason why it is even possible to apply the derived expressions for the entropy to de Sitter space is that they are agnostic about the precise background under consideration. While originally studied in the context of holography in Anti-de Sitter space, by now many works have applied this so-called island formula to other backgrounds, such as a plethora of black hole solutions \cite{Gautason:2020tmk,Hartman:2020swn,Balasubramanian:2020coy,Miyata:2021ncm,Balasubramanian:2021wgd,Chandrasekaran:2021tkb,Miyata:2021qsm,DeVuyst:2022bua,Yu:2022xlh,Murdia:2022giv} and cosmological spacetimes \cite{Chen:2020tes,Hartman:2020khs,Sybesma:2020fxg,Balasubramanian:2020xqf,Geng:2021wcq,Aalsma:2021bit,Aguilar-Gutierrez:2021bns,Langhoff:2021uct,Kames-King:2021etp,Goswami:2021ksw,Bousso:2022gth,Espindola:2022fqb,Svesko:2022txo,Levine:2022wos,Azarnia:2022kmp,Yadav:2022jib,Goswami:2022ylc}. As long as it is possible to unambiguously define a subsystem entangled with a gravitating region, the island formula applies. It has even been suggested that disconnected entanglement wedges, a feature of the island formula, are property of gravitating regions as well \cite{Bousso:2022hlz}.

Quantitatively, the fine-grained entropy of a region $R$ is given by
\beq \label{eq:island-formula}
S_{\rm gen}(R) = \text{min~} \text{ext}_i\left[\frac{A(i)}{4G_N} + S_{\rm vN}(R\cup I)\right] ~.
\eeq
Here we have to allow for the possible contribution of an island region $I$ with endpoint $i$. This expression consists of an area term $A(i)$ of the co-dimension-two surface $i$ and the von Neumann entropy of $R\cup I$. The location of $i$ is found by extremizing the generalized entropy. When an extremum is found, $i$ is known as a quantum extremal surface (QES). If there are multiple extrema, the island formula picks out the one that gives the lowest entropy. As mentioned, using the island formula to probe quantum effects in de Sitter space is perhaps one of the most exciting prospects of this program. 
However, for a variety of reasons that we will explain, applying the island formula in de Sitter space is less straightforward than in black hole spacetimes.

First, there is no non-gravitating asymptotic region in de Sitter space to which \eqref{eq:island-formula} can directly be applied. Perhaps, this is just a technical obstruction requiring a more subtle treatment of entanglement wedges in gravitating regions 

Second, constant time slices of de Sitter space are compact. This implies that, when de Sitter space is entangled with a non-gravitating disjoint auxiliary space $R$ such that the total system is pure, the dominant island covers the entirety of de Sitter space. Owning to its compactness, this surface has a vanishing area which leads to a vanishing generalized entropy when $R$ covers the entire auxiliary space \cite{Almheiri:2019hni,Balasubramanian:2020xqf,Shaghoulian:2021cef}. This is the correct answer because the total system is pure, but it does not teach us much about the constraints finite de Sitter entropy puts on semi-classical physics. Even when $R$ covers only part of the non-gravitating region, the resulting generalized entropy is just the von Neumann entropy of $R$ without receiving any contribution from de Sitter space \cite{Mahajan21}.

Third, the bifurcation surface in de Sitter space is a minimax surface, meaning that it is a quantum extremal surface that is minimal in time and maximal in space. This should be contrasted with the bifurcation surface of a black hole, which is maximin (maximal in time, minimal in space).\footnote{Timelike separated islands have also been considered in de Sitter space, based on analytic continuation from Euclidean AdS \cite{Chen:2020tes}. A possible interpretation of timelike entropy has been suggested recently in \cite{Doi:2022iyj}.} The consequence of this difference is that, when one computes the fine-grained entropy of a subregion in de Sitter space defined by freezing gravity in a region that is small with respect to the horizon, the resulting QES displays puzzling behavior. For example, in \cite{Sybesma:2020fxg} this led to an island moving back in time, and in \cite{Shaghoulian:2021cef} it was shown that a minimax QES leads to a violation of entanglement wedge nesting.

At the same time, it has been shown that conventional (maximin) quantum extremal surfaces do exist when considering a non-equilibrium state that corresponds to an observer that collects radiation in a static patch \cite{Aalsma:2021bit}, albeit at the price of inducing large backreaction. In fact, backreaction seems to be a crucial ingredient if one wants to decode information from the cosmological horizon \cite{Aalsma:2021kle}. This is perhaps not so surprising, as the finite volume of the static patch implies that any collected radiation will induce backreaction. 

In this paper, we take a closer look at the appearance of a maximin QES in a backreacted de Sitter geometry, continuing previous work \cite{Aalsma:2021bit}. Our main focus is understanding how information can be decoded from the Gibbons-Hawking radiation, without inducing catastrophic backreaction. We do so by constructing a cosmological model in two-dimensional gravity with a domain wall that interpolates between a de Sitter and a (gravitating) Rindler geometry. This has the effect of introducing asymptotic regions where gravity decouples such that the island formula can be applied unambiguously. This asymptotic region gives us an outside perspective of the cosmological horizon, evading some of the thorny issues of defining a subsystem in de Sitter space where the island formula can be applied to. The use of auxiliary subsystems has recently proven to be useful to study de Sitter thermodynamics \cite{Svesko:2022txo,Banihashemi:2022htw}.

Although the resulting geometry is quite a drastic departure from the geometry of pure two-dimensional de Sitter space, we argue that this setup still allows us to get information about de Sitter space by computing the entropy of radiation being emitted by the cosmological horizon. We show that by entangling the Rindler region with de Sitter space in the Bunch-Davies state, the entropy of radiation saturates at the de Sitter entropy by the presence of a conventional maximin QES. However, this island can be associated to the Rindler region and does not directly teach us about de Sitter space. This situation changes when we break the thermal equilibrium. By briefly doing so, a de Sitter island appears that allows information recovery from the cosmological horizon in a controlled fashion, reproducing Page-like behavior.

The rest of this article is organized as follows. In Section \ref{eq:2dModels} we present our two-dimensional gravity model with a domain wall that interpolates between de Sitter space and a Rindler wedge. We show how the junction conditions are solved and discuss the various quantum states. Then, in Section \ref{sec:entropy} we compute the entropy using the island formula and discuss different scenarios of information recovery. Lastly, we discuss our results and outline some future directions in Section \ref{sec:discussion}.

\section{Two-Dimensional Dilatonic Gravity Models} \label{eq:2dModels}
In this section, we consider two different two-dimensional gravity models. The first one is JT gravity \cite{Teitelboim:1983ux,Jackiw:1984je} that gives rise to a dS$_2$ solution. The second one is the CGHS model \cite{Callan:1992rs}, which allows for flat space solutions. We show that we can introduce a domain wall that glues these models together. By solving the two-dimensional analog of the Israel junction conditions, we show that this leads to a spacetime that has asymptotic regions where gravity decouples. We furthermore comment on the higher-dimensional interpretation of these separate models.

\subsection{De Sitter and Rindler Solutions}
We will consider two different two-dimensional gravity solutions coupled to a dilaton and conformal matter. First, we study JT gravity on a de Sitter background
\beq
I_{\rm JT} = \frac{\Phi_0}{2\kappa^2}\int \rmd^2x\sqrt{-g}R+\frac1{2\kappa^2}\int \rmd^2x\sqrt{-g}\Phi(R-2/\ell^2) + I_{\rm BDY} + I_{\rm CFT} ~.\label{eq:actionJTdS}
\eeq
Here $\kappa^2=8\pi G_N$. The first term is a topological term proportional to a constant $\Phi_0$ that we take to be large and positive. The second term contains the dynamical dilaton $\Phi$ and the last two terms describe boundary terms and the coupling to conformal matter. The equations of motion are given by
\beq
\bal
 - \nabla_a\nabla_b\Phi + g_{ab}\square\Phi + \frac{\Phi}{\ell^2}g_{ab} - \kappa^2 \langle
 T_{ab}
 \rangle&= 0 ~,\\
R-2/\ell^2 &= 0 ~,\label{eq:EOMJTdS}
\eal
\eeq
where $a,b$ run over space and time.
The equation of motion for the dilaton sets the Ricci scalar with constant positive curvature. 
The conformal matter has a stress tensor $\langle T_{ab}\rangle$ whose vacuum expectation value is determined by the metric and quantum state under consideration.
In particular, in two dimensions, the conformal anomaly takes the form 
\begin{equation}
    \langle
        T^{a}{_a}
    \rangle
    =
    \frac{c}{12\pi \ell^{2}}
    \,,
\end{equation}
where $c$ is the central charge of the conformal matter. In this work, we consider $c\gg1$ such that we do not have to take into account other semi-classical corrections.

Following \cite{Christensen:1977jc}, we use the expression for the conformal anomaly above combined with the conservation of the expectation value of the stress tensor to derive
\begin{equation}\label{eq:semiemt}
	\langle T_{\pm\pm}(x^\pm)\rangle
	=
	\frac{c}{12\pi}
	\left[
		\partial_{\pm}^{2}\omega
		-
		(\partial_{\pm}\omega)^{2}
	\right]
	+
	\langle:T_{\pm\pm}(x^{\pm}):\rangle
	\,,
\end{equation}
\begin{equation}
	\langle T_{+-}\rangle
	=
	-\frac{c}{12\pi}\partial_{+}\partial_{-}
	\omega
	\,,
\end{equation}
where we adopted conformal gauge, $\rmd s^{2}=-{e}^{2\omega}\rmd x^{+}\rmd x^{-}$. The object $\langle:T_{\pm\pm}(x^{\pm}):\rangle$ is the normal-ordered stress tensor which quantifies the flux that an observer using coordinates $x^{\pm}$ measures. It is important to stress that under coordinate transformation $x^{\pm}(y):=x^{\pm}(y^{\pm})$ it does not quite transform as a tensor, but picks up an anomalous piece,
\begin{equation}\label{eq:trafolaw}
	\langle:T_{\pm\pm}(x^{\pm}(y)):\rangle
	=
    (\left.x^{\pm}\right.'(y))^{-2}
	\langle:T_{\pm\pm}(y):\rangle
	+
    \frac{c}{24\pi}
    (\left.x^{\pm}\right.'(y))^{-2}
	\{x^{\pm}(y),y^{\pm}\}
	\,,
\end{equation}
where the curly brackets denote the Schwarzian derivative
\begin{equation}
        \{x^{\pm}(y),y^{\pm}\}
        =
		\frac{\left.x^{\pm}\right.'''(y)}{\left.x^{\pm}\right.'(y)}
		-
		\frac{3}{2}
		\left(
			\frac{\left.x^{\pm}\right.''(y)}{\left.x^{\pm}\right.'(y)}
		\right)^{2}
        \,.
\end{equation}
The anomalous transformation of $\langle:T_{\pm\pm}:\rangle$ allows $\langle T_{\pm\pm}\rangle$ in \eqref{eq:semiemt} to transform as a tensor, as it compensates a term coming from the fact that $\omega$ does not transform as a scalar under coordinate transformations. 

In what follows, we will discuss different quantum vacuums that are naturally defined in different coordinate systems. We first consider a light cone version of the static coordinates, $\sigma^{\pm}=t\pm \rho$, where $t$ is the static patch time coordinates and $\rho$ is the radial tortoise coordinate. The corresponding metric in static coordinates is given by
\begin{equation}
    \rmd s^{2}
    =
    -\text{sech}{^2}\left(
        \frac{\sigma^{+}-\sigma^{-}}{2\ell}
    \right)
    \rmd\sigma^{+}\rmd\sigma^{-}
    \,.
\end{equation}
We first consider the conformal matter to be in the natural vacuum state, which is the Bunch-Davies vacuum. 

A static observer, who uses $t=(\sigma^{+}+\sigma^{-})/2$ to perform normal ordering, measures both left and right moving flux corresponding to black body radiation at the Gibbons-Hawking temperature $T_{\text{GH}}=(2\pi\ell)^{-1}$. This implies that the normal-ordered stress tensor takes the form
\begin{equation}\label{eq:thermal-stress}
    \text{Bunch-Davies Vacuum:}
    \quad
    \langle:T_{\pm\pm}(\sigma^{\pm}):\rangle
    \,
    =
    \frac{\pi c}{12}T^{2}_{\text{GH}}
    =
    \frac{c}{48\pi \ell^{2}}
    \,.
\end{equation}
Using \eqref{eq:semiemt}, we see that this corresponds to
\begin{equation}
    \langle T_{\pm\pm}(\sigma^\pm)\rangle
    =
    0
    \,.
\end{equation}
The static patch coordinates $\sigma^{\pm}$ cover only the static patch or one-quarter of the conformal diagram. For computations, we find it convenient to extend the valid range of the coordinates. To do so we introduce Kruskal coordinates $x^{\pm}$, which can be related to static coordinates $\sigma^{\pm}$ via
\begin{equation}\label{eq:static-coord}
    x^{\pm}
    =
    \pm\ell {e}^{\pm\sigma^{\pm}/\ell}
    \,.
\end{equation}
Using the anomalous transformation law \eqref{eq:trafolaw}, we find
\begin{equation}
    \langle T_{\pm\pm}(x^{\pm})\rangle
    =
    0
    \,,
\end{equation}
where we remind the reader that this is the object that sources the Einstein field equations and transforms them as a tensor. 
To compute the flux an observer measures in the Bunch-Davies vacuum using the Kruskal coordinates $x^{\pm}$, instead of static coordinates $\sigma^{\pm}$, we need to compute the normal-ordered stress tensor using the anomalous transformation law \eqref{eq:trafolaw}.  
This leads to
\begin{equation}
    \text{Bunch-Davies Vacuum:}
    \quad
    \langle:T_{\pm\pm}(x^{\pm}):\rangle \,  
    =
    0
    \,.
\end{equation}
This implies that an observer using $x^{\pm}$ coordinates for normal ordering does not experience any temperature in the Bunch-Davies vacuum.

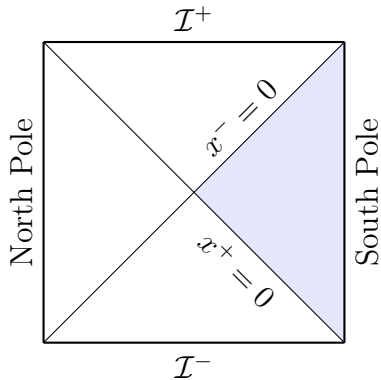
\begin{figure}[t]
\centering
\begin{tikzpicture}

\node[isosceles triangle,
    isosceles triangle apex angle=90,
    fill=blue!10,
    rotate=180,
    minimum size =2cm] (T90)at (3.2,2){};

\draw[thick] (0,0) -- (4,0) node[pos=.5,below]{${\cal I}^-$};
\draw[thick] (0,0) -- (0,4) node[pos=.5,above,rotate=90]{North Pole};
\draw[thick] (0,4) -- (4,4) node[pos=.5,above]{${\cal I}^+$};
\draw[thick] (4,0) -- (4,4) node[pos=.5,below,rotate=90]{South Pole};

\draw (0,0) -- (4,4) node[pos=.7,above,rotate=45]{$x^-=0$};
\draw (0,4) -- (4,0) node[pos=.7,below,rotate=-45]{$x^+=0$};
\end{tikzpicture}
\caption{Penrose diagram of de Sitter space with conformal matter in the Bunch-Davies state. The shaded region corresponds to the South Pole static patch. Here, we did not include the region corresponding to $\Phi< 0$ because that will be absent from our final patched geometry.}
\label{fig:Pen-dS}
\end{figure}

With the metric and stress tensor in hand, we can solve the equation of motion for the dilaton. A general solution to the JT equations of motion in the Bunch-Davies vacuum is
\beq \label{eq:gen-dil-dS}
\Phi = \frac{c\kappa^2}{24\pi} +\phi_0\frac{\ell^2+x^+x^-}{\ell^2-x^+x^-} + \frac{\ell}{\ell^2-x^+x^-}\left(C_1x^++C_2 x^-\right) ~.
\eeq
Here $\phi_0$ is a constant and $C_1,C_2$ are constants that multiply time-dependent contributions in terms of $\sigma^{\pm}$ coordinates. We require $\Phi_{0}\gg\phi_{0}\gg c\gg1$ to suppress corrections within the semi-classical approximation. For the Bunch-Davies vacuum, we are mostly interested in time-independent solutions, so we will set $C_1=C_2=0$. The Penrose diagram is given by Figure \ref{fig:Pen-dS}. Evaluating the dilaton at the horizon, the entropy is found to be
\beq
S_{\rm dS} = \frac{2\pi}{\kappa^2}\left(\Phi_0+\Phi\right)_{x^-=0} = \frac{2\pi}{\kappa^2}(\Phi_0+\phi_0) + \frac c{12} ~.
\eeq
The North and South pole are given by $x^+x^-=-\ell^2$ and ${\cal I}^\pm$ are described by $x^+x^-=\ell^2$. At ${\cal I}^{\pm}$, $\Phi\to\infty$ and at the poles it takes the constant value $\Phi = c\kappa^2/(24\pi)$.

The second model we study is the CGHS model, whose actions can be brought into the following form using a Weyl transformation
\beq \label{eq:actionJTflat}
I_{\rm CGHS} =  
 \frac{\Phi_0}{2\kappa^2}\int \rmd^2x\sqrt{-g}R +\frac1{2\kappa^2}\int \rmd^2x\sqrt{-g}(\Phi R+4\lambda^2) + I_{\rm BDY} + I_{\rm CFT} ~.
\eeq
As we will elaborate upon a bit more later on, the gravitational part of this model can be obtained from the dimensional reduction of a higher-dimensional theory, in which $\lambda$ can be interpreted as a cosmological constant. The first term corresponds to a topological term that does not contribute to the equations of motion. However, it can possibly contribute to the (Wald) entropy \cite{Pedraza:2021cvx,Pedraza:2021ssc}. As we explain more in detail, later on, we view the CGHS model as the dimensional reduction of the near-horizon geometry of a (non-extremal) black hole. In this case, $\Phi_0$ parametrizes its entropy similar to JT gravity. \footnote{To see that the CGHS model can contain a topological term, see e.g. \cite{Godet:2021cdl}.}

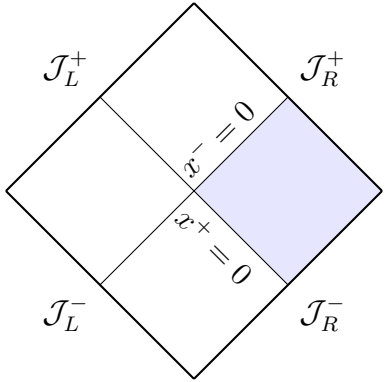
\begin{figure}[t!]
\centering
\begin{tikzpicture}

\fill [blue!10,rotate around={-45:(2.5,0)}] (2.5,0) rectangle (4.25,1.75);

\draw[thick] (0,0) -- (2.5,2.5) node[pos=.5,above left]{$\mathcal{J}_L^+$};
\draw[thick] (2.5,2.5) -- (5,0) node[pos=.5,above right]{$\mathcal{J}_R^+$};
\draw[thick] (0,0) -- (2.5,-2.5) node[pos=.5,below left]{$\mathcal{J}_L^-$};
\draw[thick] (2.5,-2.5) -- (5,0) node[pos=.5,below right]{$\mathcal{J}_R^-$};;

\draw (1.25,-1.25) -- (3.75,1.25) node[pos=.7,rotate=45,above]{$x^-=0$};
\draw (1.25,1.25) -- (3.75,-1.25) node[pos=.7,rotate=-45,below]{$x^+=0$};

\end{tikzpicture}
\caption{Penrose diagram of flat space. The shaded region corresponds to the right Rindler wedge.}
\label{fig:Pen-flat}
\end{figure}

The corresponding equations of motion are given by
\beq
\bal
-\nabla_a\nabla_b\Phi + g_{ab}\square\Phi -2\lambda^2g_{ab} - \kappa^2\langle T_{ab}\rangle &=0 ~, \\ 
R &= 0  ~.
\eal\label{eq:EOMJTflat}
\eeq
Clearly, this model describes flat space and we can introduce the following coordinates
\beq
\rmd s^2 = - \rmd x^+\rmd x^- ~.
\eeq
Because we are interested in coupling the de Sitter solution to flat space, it is appropriate to take the conformal matter in the Minkowski vacuum with a stress tensor given by
\beq
\text{Minkowksi Vacuum:}\qquad \langle:T_{\pm\pm}(x^\pm):\rangle\, = 0 ~,
\eeq
where $x^{\pm}=t\pm x$ brings the metric into a manifest flat form. As is well known, the Minkowksi vacuum looks thermal to an accelerating (Rindler) observer. We can now introduce Rindler coordinates with a temperature equal to the de Sitter temperature. The relationship between flat and Rindler coordinates is then again given by \eqref{eq:static-coord}. Similarly, the normal-ordered stress tensor in Rindler coordinates takes the form \eqref{eq:thermal-stress}. The dilaton solution to the CGHS equations of motion is now given by
\beq
\Phi_0+\Phi = \frac{c\kappa^2}{24\pi} + \Phi_0+\phi_0 - \lambda^2x^+x^- + D_1x^++D_2x^- ~.
\eeq
The two integration constants $D_1,D_2$ capture time-dependent contributions, so we again set those to zero: $D_1=D_2=0$. The CGHS model can be obtained by dimensional reduction at least in two ways: as the near-horizon geometry of a BTZ black hole, in which case $\lambda$ can be related to the cosmological constant, and as the near-horizon geometry of a four-dimensional near-extremal charged black hole, in which case $\lambda$ is related to the charge. In two dimensions it seems to appear as a free parameter, but we will set $\lambda = \sqrt{\phi_0}/\ell$. This ensures that within a static patch/Rindler wedge, the value of the dilaton uniquely determines whether we are in the de Sitter or Rindler region. We discuss the significance of this observation in the next subsection. The remaining constants are chosen such that the entropy matches the de Sitter entropy at the Rindler horizon. The right Rindler wedge is indicated by the shaded region in Figure \ref{fig:Pen-flat}. Within these regions of interest, the asymptotic regions are reached as
\beq
\bal
{\cal J}^{\pm}_L: \quad &x^{\mp}\to \pm\infty ~, \\
{\cal J}^\pm_R: \quad &x^\pm\to\pm\infty ~.
\eal
\eeq
The dilaton will grow towards the asymptotic regions, signaling the weakening of gravity and finally, the decoupling limit when $\Phi\to\infty$. 

\subsection{Domain Wall Between De Sitter and Rindler Space}
We now construct a two-dimensional gravity model by introducing a domain wall that interpolates between de Sitter space and a Rindler wedge. In order for such a patched spacetime to solve the equations of motion, the analogs of the Israel junction conditions in two-dimensional dilaton gravity need to be satisfied. These conditions have recently been explored in \cite{Engelhardt:2022qts} (see also \cite{Mirbabayi:2020grb} for earlier work). It is convenient to think of the dilaton solutions considered above as solutions of an Einstein equation of the form $G_{ab}-\kappa^2 T_{ab}=0$, where the Einstein tensor has different cosmological constant parameters for the JT and CGHS model. Consider:
\beq
\bal
\text{JT:} \qquad G_{ab} &=  - \nabla_a\nabla_b\Phi + g_{ab}\square\Phi + \Phi\Lambda g_{ab} ~, \\
\text{CGHS:} \qquad G_{ab} &= - \nabla_a\nabla_b\Phi + g_{ab}\square\Phi -2\lambda^2g_{ab}~. \\
\eal
\eeq
We can think of the junction as separating two different spacetimes with constants $\Lambda=1/\ell^2$ and $\lambda=\sqrt{\phi_0}/\ell$.

In what follows, we will mostly interpret our setup as a particular two-dimensional model. We will however briefly comment on its higher-dimensional interpretation. Typically when one considers a junction between two different solutions, the two spacetimes are two different solutions of the same action. Here, we are instead interested in two solutions of different actions, corresponding to JT gravity and the CGHS model. This is consistent because both actions can be obtained from dimensional reduction of a higher-dimensional solution. For example, JT gravity on a de Sitter background can be obtained from the dimensional reduction of three-dimensional empty de Sitter space \cite{Sybesma:2020fxg} or a near-Nariai black hole in four dimensions \cite{Maldacena:2019cbz}. The CGHS model arises as the near-horizon dynamics of a (non-rotating) BTZ black hole in three dimensions \cite{Godet:2021cdl} or a charged black hole in four dimensions \cite{Giddings:1992kn,Callan:1992rs}. The latter two solutions have Rindler factors in their near-horizon geometry. Our model, therefore, shares similarities with a gravastar \cite{Mazur:2001fv}, a geometry that looks like a black hole from the outside with its interior replaced by de Sitter space. Our choice of the parameter $\lambda=\sqrt{\phi_0}/\ell$ in the CGHS model ensures that we are gluing the region outside the black hole to a de Sitter interior. This interpretation also requires the constant piece of the dilaton $\Phi_0$ to be the same in the JT and CGHS model to ensure continuity of the transverse sphere. Strictly speaking, because only the near-horizon geometry is described by Rindler space the dilaton, which is the radial coordinate in the black hole geometry, should be cut off at a large value.\footnote{This can be thought of as a consequence of the absence of a decoupling limit for non-extremal black holes.}. For our purposes, it suffices to think of the asymptotic regions in the Rindler geometry as these timelike cutoff surfaces.

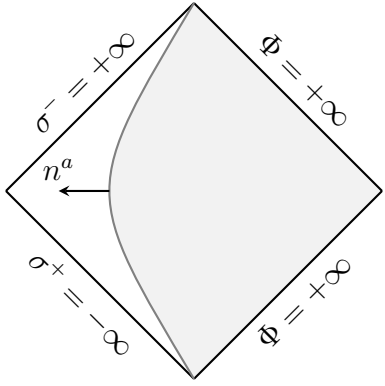
\begin{figure}[t!]
\centering
\begin{tikzpicture}

\node[isosceles triangle,
    isosceles triangle apex angle=90,
    fill=gray!10,
    minimum size =2.5cm] (T90)at (3.5,0){};

\draw[thick] (0,0) -- (2.5,2.5) node[pos=.5,rotate=45,above]{$\sigma^-=+\infty$};
\draw[thick] (2.5,2.5) -- (5,0) node[pos=.5,above,rotate=-45]{$\Phi=+\infty$};
\draw[thick] (0,0) -- (2.5,-2.5) node[pos=.5,rotate=-45,below]{$\sigma^+=-\infty$};
\draw[thick] (2.5,-2.5) -- (5,0) node[pos=.5,below,rotate=45]{$\Phi=+\infty$};

\draw[thick,color=gray,fill=gray!10] (2.5,-2.5) .. controls (1,0) .. (2.5,2.5);
\draw [thick,stealth-](.7,0) -- (1.375,0) node[pos=0,above]{$n^a$};

\end{tikzpicture}
\caption{Domain wall with normal vector $n^a$ separating part of a de Sitter static patch (white) from a Rindler region (gray). The two junction conditions determine the location of the domain wall.}
\label{fig:junction}
\end{figure}

\begin{figure}[t!]
\centering
\begin{tikzpicture}

\fill [gray!10,rotate around={-45:(4,0)}] (4,0) rectangle (6.8,2.8);
\fill [gray!10,rotate around={135:(4,0)}] (4,0) rectangle (6.8,2.8);

\draw[thick] (0,0) -- (2,2) node[pos=.5,above left]{$\mathcal{J}_L^+$};
\draw[thick] (0,0) -- (2,-2) node[pos=.5,below left]{$\mathcal{J}_L^-$};
\draw[thick] (2,2) -- (6,-2) node[pos=.7,rotate=-45,below]{$\sigma^+=-\infty$};
\draw[thick] (2,-2) -- (6,2) node[pos=.7,rotate=45,above]{$\sigma^-=+\infty$};
\draw[thick] (2,2) -- (6,2) node[pos=.5,above]{$\mathcal{I}^+$};
\draw[thick] (2,-2) -- (6,-2) node[pos=.5,below]{$\mathcal{I}^-$};
\draw[thick] (6,2) -- (8,0) node[pos=.5,above right]{$\mathcal{J}_R^+$};
\draw[thick] (6,-2) -- (8,0) node[pos=.5,below right]{$\mathcal{J}_R^-$};

\draw[thick,darkgray] (6,-2) .. controls (7.5,0) .. (6,2) node[pos=.5,right] {${\cal C}_R$};
\draw[thick,darkgray] (2,-2) .. controls (0.5,0) .. (2,2) node[pos=.5,left] {${\cal C}_L$};

\draw[red,very thick] (6.7,1) -- (8,0) node[pos=1,right]{$R_R$};
\filldraw[red] (6.7,1) circle (2.5pt);
\draw[red,very thick] (0,0) -- (1.3,1) node[pos=0,left]{$R_L$};
\filldraw[red] (1.3,1) circle (2.5pt);
\end{tikzpicture}
\caption{Penrose diagram of the glued solution where Rindler wedges (shaded gray) replace the static patches of de Sitter space. We define anchor curves ${\cal C}_{L,R}$ where $\Phi\gg\Phi_0\gg 1$ from which the de Sitter region (white) can be probed.
}
\label{fig:dS-Rind-models}
\end{figure}
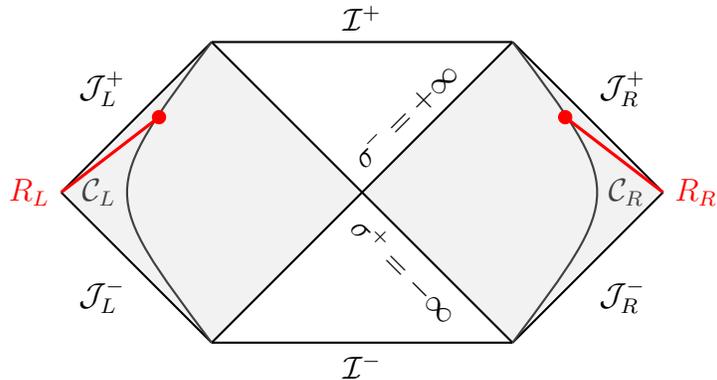

To define the domain wall between two different spacetimes, we parametrize its trajectory by $x^a(\mathfrak{t})$, where $\mathfrak{t}$ is a time coordinate on the domain wall. Its location can be parametrized by setting a scalar function to zero: $s(x^a(\mathfrak{t}))=0$. The spacetime to the left of the junction is given by $s<0$ and the spacetime to the right by $s>0$.  We denote the unit normal to the junction as $n_a$, which obeys
\beq
\dot x^an_a = 0 ~, \quad n^an_a=1 ~.
\eeq
Here the dot denotes $\frac{\rmd}{\rmd\mathfrak{t}}$. To give explicit expressions for the unit normal, we find it convenient to work in conformally flat coordinates. In the static patch of de Sitter space, this is achieved by working with the tortoise coordinate $\rho = -\ell\, \text{arctanh}(r/\ell)$. The metric is then
\beq
\rmd s^2 = \text{sech}^2(\rho/\ell)(-\rmd t^2 + \rmd \rho^2) ~.
\eeq
We are interested in a timelike junction with a left-pointing normal vector, see Figure \ref{fig:junction}.
The unit normal in the de Sitter region is given by
\beq
n_a = \frac{1}{\cosh(\rho/\ell)\sqrt{\dot t^2-\dot \rho^2}}\{\dot \rho,-\dot t\}\,.
\eeq
We now choose an embedding $\mathfrak{t} = t$. There are now two junction conditions that need to be satisfied. We use square brackets to denote the difference between the left and the right of the junction, e.g. $[A] = A(s<0) - A(s>0)$. Then, the first condition requires the dilaton to be continuous \cite{Engelhardt:2022qts},
\beq
[\Phi] = 0 ~.
\eeq
The second condition determines the trajectory of the domain wall in terms of its tension ${\cal T}$ \cite{Mirbabayi:2020grb,Engelhardt:2022qts},
\beq \label{eq:2ndjunction}
\kappa^2{\cal T} = [n^a\partial_a \Phi] ~.
\eeq
We are interested in a tensionless domain wall separating the two spacetimes such that radiation can move freely through it. We, therefore, impose
\beq
[\Phi]_{s=0}=[n^a\partial_a\Phi]_{s=0} = 0 ~.
\eeq
In terms of static coordinates,
\beq
x^\pm = \pm \ell e^{\pm(t\pm \rho)/\ell} ~,
\eeq
the relevant dilaton solutions are given by
\beq
\bal
\text{de Sitter:}\quad \Phi &= \frac{c\kappa^2}{24\pi} -\phi_0\,\tanh(\rho/\ell) ~, \\
\text{Rindler:}\quad \Phi &=  \frac{c\kappa^2}{24\pi} + \phi_0(1 + e^{2\rho/\ell}) ~.
\eal
\eeq
In these coordinates, we find that the first junction condition is satisfied when $\rho\to-\infty$, which corresponds to the horizon. Instead of explicitly solving for the trajectory of the brane using the second junction condition, we simply look for a static solution where $\dot\rho=0$. Imposing this condition, we indeed find using \eqref{eq:2ndjunction} that this corresponds to a tensionless domain wall located at $\rho\to-\infty$.\footnote{Close to the horizon, the de Sitter normal vector behaves as $n^a = -\frac12\{0,e^{-\rho/\ell}\}$. To match onto the normal vector in Rindler space, we, therefore, need to rescale the radial coordinate to remove the factor $1/2$.} The resulting geometry is shown in Figure \ref{fig:dS-Rind-models}. By performing this gluing procedure, we already broke the full de Sitter isometries down to those preserved by a static patch. In fact, this is necessary for a static observer to recover information from behind the cosmological horizon \cite{Aalsma:2021bit}.

Because this model has asymptotic regions ${\cal J}_{L,R}^{\pm}$ where the dilaton diverges positively, we can define two asymptotic curves ${\cal C}_{L,R}$ (indicated in Figure \ref{fig:dS-Rind-models}) where gravity becomes weak. We can define these curves in a covariant manner by setting the dilaton equal to a constant and large value: $\Phi=L/\ell\gg \Phi_{0}\gg1$, where $L$ is the largest length scale of the problem. In terms of Rindler coordinates, this corresponds to taking $\rho/\ell \gg 1$. By taking $\rho$ sufficiently large, gravity can be made arbitrarily weak and we can compute the generalized entropy of $R$ using the island formula.

\subsection{Breaking the Thermal Equilibrium}
In the above model, the Rindler wedges that replace the static patches remain in thermal equilibrium with de Sitter space. However, we can also break this thermal equilibrium by, from the perspective of the right Rindler observer, removing the left-moving component of the radiation effectively turning off the heat bath. This way, the observer will only measure the radiation that is originating from the de Sitter region, in principle allowing them to recover information from behind the horizon.

Removing the left-moving part of the radiation can be achieved by setting the corresponding normal-ordered stress tensor to zero \cite{Aalsma:2019rpt,Aalsma:2021bit}. However, as discussed in \cite{Aalsma:2021bit,Kames-King:2021etp} this eventually leads to large and catastrophic backreaction due to a permanent flux of energy. Therefore, instead of considering a solution that breaks the thermal equilibrium eternally, we will consider a setup where the equilibrium is only broken during an interval $x^+_A\leq x^+\leq x^+_B$. We refer to this interval as a ``pulse'' (as opposed to a sharply localized shockwave). This does not only have the advantage that we can keep control over the magnitude of backreaction, but also provides us with a dynamical way of setting up the non-equilibrium state, regulating the infinite energy present in the eternal non-equilibrium state \cite{Aalsma:2021bit}.

We will consider the expressions for the diagonal components of the stress tensor in Kruskal coordinates as follows
\beq\label{eq:stress tensor non equil}
\bal
\langle T_{++}(x^+)\rangle &= -\frac{c}{48\pi}\left[\frac{\theta(x^+-x^+_A)-\theta(x^+-x^+_B)}{(x^+)^2} + \frac{\delta(x^+-x^+_A)}{x^+_A}-\frac{\delta(x^+-x^+_B)}{x^+_B}\right] ~, \\
\langle T_{--}(x^-)\rangle &= 0  ~.
\eal
\eeq
The form of the stress tensor needs some explanation. The terms proportional to the Heaviside theta function turn off the left-moving radiation during $x^+_A\leq x^+\leq x^+_B$. In addition, we also included shockwaves at the location where the pulse is turned on and off, which are given by the terms proportional to the delta functions. These shockwave contributions ensure that there are no discontinuities in the (integrated) Killing energy when crossing the pulse.

During the pulse $x^+_A\leq x^+\leq x^+_B$ the normal-ordered stress tensor is
\beq
\bal \label{eq:NonEqState}
\langle:T_{++}(\sigma^+):\rangle &= \frac{c}{48\pi \ell}\left({e}^{2(\sigma^+-\sigma^+_A)/\ell}\delta(\sigma^+-\sigma^+_A)-{e}^{2(\sigma^+-\sigma^+_B)/\ell}\delta(\sigma^+-\sigma^+_B)\right) ~,\\
\langle:T_{--}(\sigma^-):\rangle &= \frac{c}{48\pi\ell^2} ~.
\eal
\eeq
Because the $++$ component of the stress tensor parametrizes radiation along a constant $\sigma^+$ slice, this indeed shows that there is no left-moving radiation during the pulse, except at the shocks at $x^+=x^+_A$ and $x^+=x^+_B$.

Of course, this new stress tensor will induce backreaction and modify the dilaton solutions. Reconsidering the equations of motion, we now find that the dilaton solutions are given by
\begin{equation}\label{eq:domaindSRindler}
    \Phi=
    \begin{cases}
        \frac{c\kappa^2}{24\pi} +\phi_0\frac{\ell^2+x^+x^-}{\ell^2-x^+x^-} - \frac{c\kappa^2}{48\pi}\left(H_{\rm dS}(x^+_A)-H_{\rm dS}(x^+_B)\right) ~,
        \\
        \frac{c\kappa^2}{24\pi} + \phi_0(1 - \ell^{-2}x^+x^-) - \frac{c\kappa^2}{48\pi}\left(H_{\rm R}(x^+_A)-H_{\rm R}(x^+_B)\right) ~,
    \end{cases}
\end{equation}
where the top line denotes the de Sitter region and the bottom line denotes the Rindler region.
For convenience, we defined the functions
\beq
\bal
H_{\rm dS}(x^+_j) &= \left[-2x^-\frac{x^+-x^+_j}{\ell^2-x^+x^-} + \frac{\ell^2+x^+x^-}{\ell^2-x^+x^-}\log\left(x^+/x^+_j\right)\right]\theta(x^+-x^+_j) ~, \\
H_{\rm R}(x^+_j) &= \log\left(x^+/x^+_j\right)\theta(x^+-x^+_j) ~,
\eal
\eeq
with the index $j\equiv(A,B)$. We note that the dilaton solution in de Sitter space during $x^+_A<x^+<x^+_B$ does not precisely reduce to the non-equilibrium dilaton solution considered in \cite{Aalsma:2021bit}. The reason for this difference can be attributed to the different initial conditions that we are considering in the present work. Here, we start out from the Bunch-Davies state and dynamically transition to the non-equilibrium state. This fixes the integration constants in the general dilaton solution \eqref{eq:gen-dil-dS}. In contrast, \cite{Aalsma:2021bit} considered an eternal non-equilibrium state, which amounts to a different choice of integration constants. Importantly, the modified dilaton solution still diverges positively at the asymptotic regions such that ${\cal C}_{L,R}$ corresponds to regions of weak gravity.

We now have to reassess the junction conditions in the regime $x^+_A<x^+<x^+_B$ between the Rindler and de Sitter region with the new dilaton solutions. First, we note that the first junction condition is still satisfied at the future horizon $x^-=0$. The new dilaton solution does however induce a change in the stress tensor of the junction at $x^+>x^+_A$. Imposing stationarity of the brane ($\dot\rho=0$) we find at the horizon the following tension
\beq
\bal
x^+_A<x^+<x^+_B: \quad {\cal T} &= - \frac{c }{48\pi\ell^2}x^+_Ae^{-t/\ell} ~, \\
x^+>x^+_B>x^+_A: \quad {\cal T} &= - \frac{c }{48\pi\ell^2}(x^+_A-x^+_B)e^{-t/\ell} ~.\\
\eal
\eeq
This shows that a tensionless brane (${\cal T}=0$) is no longer stationary when the pulse is applied. However, the tension required to be stationary decays exponentially fast at a rate that is faster than all other timescales of the problem. Thus, what happens is that the tensionless brane moves a bit when the pulse is applied but settles down again at $x^-=0$ exponentially fast.

We can understand the effect of breaking the thermal equilibrium on the geometry by analyzing the behavior of the dilaton. First, we consider the location of the apparent horizons. Before the thermal equilibrium is broken ($x^+<x^+_A$), the critical points of the dilaton $\partial_\pm \Phi =0$ (which determine the location of the apparent horizon) are at $x^\pm=0$. Then, when the left-moving radiation is turned off between $x^+_A<x^+<x^+_B$ there is a net flux of energy through the horizons. Importantly, the positivity/negativity of this energy is different in the de Sitter and Rinder regions. A positive net flux of energy as measured by a Rindler or de Sitter observer is
\beq
{\cal F}_{\rm Rind/dS} = \langle:T_{\pm\pm}(\sigma^\pm):\rangle - \langle:T_{\mp\mp}(\sigma^\mp):\rangle ~,
\eeq
where the upper sign corresponds to Rindler space and the lower sign to de Sitter space. The difference in sign can be attributed to the opposite sign in the first law of the horizons, see \cite{Banihashemi:2022htw} for a recent discussion. Thus, the non-equilibrium state \eqref{eq:NonEqState} contains a positive flux of Killing energy with respect to a de Sitter observer, but a negative one from the perspective of a Rindler observer. 

What both observers agree upon is that the entropy proportional to the size of the dilaton decreases. This can be seen by evaluating the dilaton at the future horizon $x^-=0$:
\beq
\left.{\frac{2\pi}{\kappa^2}(\Phi+\Phi_0)}\right|_{x^-=0} = \frac{c}{12}+\frac{2\pi}{\kappa^2}(\Phi_0+\phi_0) - \frac{c}{24}\log(x^+/x^+_A) ~.
\eeq
This process of entropy depletion can continue until there's no more entropy left, which happens at
\beq
x_{\rm end}^+=x^+_A \exp\left(2+\frac{48\pi (\Phi_0+\phi_0)}{c\kappa^2}\right) ~.\label{eq:decay to end}
\eeq
Thus, we have to make sure we stay in the regime
\beq \label{eq:BackLimit}
x^+_B \ll x^+_{\rm end} ~,
\eeq
in order to avoid strong gravitational coupling. Next, we will compute the generalized entropy and see how we can decode information from the Gibbons-Hawking radiation while remaining in a regime of control.

\section{Entropy Computations} \label{sec:entropy}
We now look for islands by extremizing the generalized entropy of $R$ as defined in Figure \ref{fig:dS-Rind-models}. Because the Rindler wedges purify de Sitter regions behind the domain wall, we make use of the fact that the von Neumann part of the generalized entropy can be computed from the region complementary to $R\cup I$. Then, using the notation of Figure \ref{fig:IslandSetup}, we have
\beq
S(R) = \text{min~} \text{ext}_{i,\tilde i}\left[\frac{2\pi}{\kappa^2}\left(\Phi_0+\Phi(x^+_i,x^-_i)+(i\leftrightarrow\tilde i)\right) + S_{\rm vN}((R\cup I)^c) \right]~,
\eeq
where the von Neumann entropy of an interval $[j,k]$ is
\beq
S_{\rm vN}([j,k])= \frac c6 \log\left[\frac{(y^+_j-y^+_k)(y^+_j-y^+_k)}{\Omega(y^+_j,y^-_j)\Omega(y^+_k,y^-_k)\epsilon_j\epsilon_k}\right] ~.
\eeq
The factor of $\epsilon_j\epsilon_k$ denotes the usual short-distance cutoffs of the von Neumann entropy and the factors $\Omega(x^+,x^-)$ are the conformal factors of a metric of the form
\beq
\rmd s^2 =- \Omega(y^+,y^-)^{-2}\rmd y^+\rmd y^- ~.
\eeq
In the two-dimensional semi-classical gravity setting, the expression above for the von Neumann entropy was first obtained in \cite{Fiola:1994ir}. In \cite{Almheiri:2019hni} this expression was derived using double holography and in \cite{Pedraza:2021cvx} it was shown that the entire generalized entropy can be found by computing the Wald entropy. The $y^\pm$ coordinates are the ones in which the stress tensor takes its vacuum form
\beq
\langle :T_{\pm\pm}(y^\pm):\rangle = 0 ~.
\eeq
The precise form of the conformal factors and vacuum coordinates will be different depending on which part of the spacetime they are evaluated on. In the region before and after the pulse ($x^+<x_A^+$ and $x^+>x_B^+$) the conformal factors expressed in Kruskal coordinates are:
\beq
\bal
\Omega_{\rm dS}(x^+,x^-)^{-2} &= \frac{4\ell^4}{(\ell^2-x^+x^-)^2} ~,\\
\Omega_{\rm R}(x^+,x^-)^{-2} &= 1 ~.
\eal
\eeq
The vacuum coordinates are then simply
\beq
y^\pm = x^\pm ~.
\eeq
However, during the pulse ($x^+_A<x^+<x^+_B$) we have
\beq
\bal
\Omega_{\rm dS}(x^+,x^-)^{-2} &= \frac{4\ell^3 x^+}{(\ell^2-x^+x^-)^2} ~,\\
\Omega_{\rm R}(x^+,x^-)^{-2} &= \frac{x^+}{\ell} ~,
\eal
\eeq
and
\beq
y^+ = \ell \log(x^+/\ell) ~, \quad y^- = x^- ~.
\eeq
Note that due to the shocks located at $x^+_A$ and $x^+_B$ these are only the vacuum coordinates away from these points. We now have all the necessary ingredients to evaluate the generalized entropy.

\subsection{Searching for Islands}
\begin{figure}[t!]
\centering
\begin{tikzpicture}

\fill [gray!10,rotate around={-45:(4,0)}] (4,0) rectangle (6.8,2.8);
\fill [gray!10,rotate around={135:(4,0)}] (4,0) rectangle (6.8,2.8);

\draw[thick] (0,0) -- (2,2);
\draw[thick] (0,0) -- (2,-2);
\draw[thick] (2,2) -- (6,-2);
\draw[thick] (2,-2) -- (6,2);
\draw[thick] (2,2) -- (6,2);
\draw[thick] (2,-2) -- (6,-2);
\draw[thick] (6,2) -- (8,0);
\draw[thick] (6,-2) -- (8,0);

\draw[thick,darkgray] (6,-2) .. controls (7.5,0) .. (6,2);
\draw[thick,darkgray] (2,-2) .. controls (0.5,0) .. (2,2);

\draw[red,very thick] (6.7,1) -- (8,0) node[pos=1,right] {$R_R$};
\filldraw[red] (6.7,1) circle (2.5pt) node[left] {$r$};
\draw[red,very thick] (0,0) -- (1.3,1) node[pos=0,left] {$R_L$};
\filldraw[red] (1.3,1) circle (2.5pt) node[right] {$\tilde r$};

\draw[blue] (3,2) -- (6.5,-1.5) node[right] {$x^+_A$};
\draw[blue] (4,2) -- (7,-1) node[right] {$x^+_B$};

\draw[red, very thick] (2.6,1.2) -- (4.6,1.2) node[pos=.5,above]{$I$};
\filldraw[red] (2.6,1.2) circle (2.5pt) node[below] {$\tilde i$};
\filldraw[red] (4.6,1.2) circle (2.5pt) node[below] {$i$};

\end{tikzpicture}
\caption{Computing the generalized entropy of $R$, we allow for the contribution of an island region $I$. The right endpoint of the radiation region $x^\pm_r$ may reside before, during, or after the pulse. The same holds true for the island's endpoint $x^\pm_i$.}
\label{fig:IslandSetup}
\end{figure}
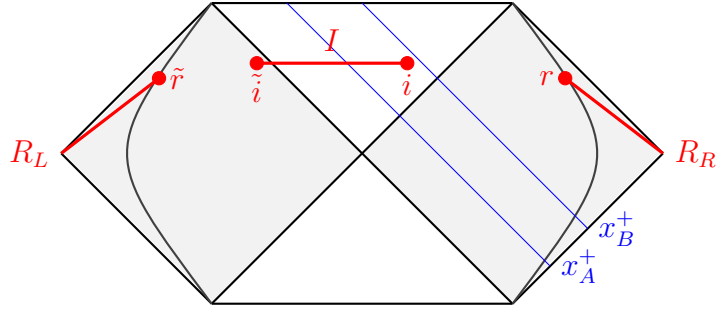

We will now extremize the generalized entropy separately for the three situations that the right observer is located before, during, and after the pulse, see Figure \ref{fig:IslandSetup}.

\subsubsection*{Trivial Island}
First, we consider a trivial or vanishing island. This is always a possible contribution and in this case, the generalized entropy is just given by the von Neumann entropy of $R=R_L \cup R_R$. Because we are working with a pure state, we can compute the entropy of $R$ from its complement. We thus have
\beq
S_{\rm vN}(R) = \frac c6\log\left[-\frac{(x^+_r-x^+_{\tilde r})(x^-_r-x^-_{\tilde r})}{\epsilon_r\epsilon_{\tilde r}}\right] ~.
\eeq
We now want to express this in Rindler coordinates, given by
\beq
\bal
\sigma^\pm_r &= \pm \ell\log(\pm x^\pm_r/\ell)~,\\
\sigma^\pm_{\tilde r} &= \mp \ell\log(\mp x^\pm_{\tilde r}/\ell) ~,
\eal
\eeq
where
\beq
\bal
\sigma^\pm_r &= t_r \pm \rho_r~,\\
\sigma^\pm_{\tilde r} &= t_{\tilde r} \mp \rho_{\tilde r} ~.
\eal
\eeq
Importantly, $t_{r,\tilde r}$ are defined such that both times flow upwards. This is necessary to break the time-translation isometry of the vacuum state that evolves time upwards in the right region and downwards in the left region. It is natural to express the UV-cutoffs also in Rindler distance. This results in the transformation
\beq \label{UVtrafo}
\epsilon_r \to \sqrt{(x_r^+)'(x^-_r)'}\,\epsilon_r ~,
\eeq
where the prime denotes a derivative with respect to $\sigma_r^\pm$ and a similar expression holds for $\epsilon_{\tilde r}$. Taking this change into account, we now consider $t _r=t_{\tilde r}$, $\rho_r=\rho_{\tilde r}$ and $\epsilon_r=\epsilon_{\tilde r}$ to find
\beq
S_{\rm vN}(R) = \frac c3\log\left[\frac{2\ell}{\epsilon_r}\cosh(t_r/\ell)\right]~.
\eeq
At $t_r/\ell\gg1$ we find
\beq
S_{\rm vN}(R) \simeq \frac c{3\ell}t_r + \frac c3\log(\ell/\epsilon_r)~,
\eeq
which grows linearly in time, eventually exceeding the de Sitter entropy. We, therefore, expect a non-trivial island to cut off this growth in order to avoid an information paradox. 

\subsubsection*{Before the Pulse}
More interesting are the non-trivial islands. First, we consider the possibility of an island in the Rindler region. In this case, we need to compute the von Neumann entropy of two disjoint intervals. We  assume that these two regions are sufficiently far separated from each other, such that we can use an OPE approximation in which the entropy is given by the sum of the separate intervals \cite{Calabrese:2009ez}. Concretely, for two intervals $[x_1,x_2]$ and $[x_3,x_4]$ we require the cross-ratio to be small:
\beq \label{eq:OPElimit}
\left|\frac{x^2_{12}x^2_{34}}{x^2_{13}x^2_{24}}\right| \ll 1 ~,
\eeq
where we defined $x_{ab}^2 = (x^+_a-x^+_b)(x^-_a-x^-_b)$. When considering an island, we will check that this condition is satisfied.

Before the pulse, the setup is left-right symmetric such that the entropy can be completely expressed in terms of the right region by exchanging $x^\pm \leftrightarrow x^\mp$. 
Assuming the existence of an island in the Rindler region in the OPE limit, the entropy is given by
\beq
S(R) = \text{min~}\text{ext}_i\left[ \frac{c}{6} + \frac{4\pi}{\kappa^2}\left(\Phi_0+\phi_0(1-\ell^{-2}x_i^+x_i^-)\right) + \frac c3\log\left[\frac{(x^+_r-x^+_i)(x^-_i-x^-_r)}{\epsilon_i\epsilon_r}\right] \right]~.
\eeq
To find an island, we need to extremize this expression. Taking derivatives with respect to $x^\pm$ and taking the limit $|x^+_rx^-_r|\gg \ell^2$ (corresponding to the asymptotic curves ${\cal C}_{L,R}$) we find the following island, expressed in Rindler coordinates,
\beq \label{eq:islandRindlerBeforePulse}
x_i^\pm = \pm \frac{c\ell\kappa^2}{12\pi\phi_0}e^{\pm(t_r\mp\rho_r)/\ell} ~.
\eeq
In the late-time limit, the island approaches the future horizon from below, such that it always lies in the Rindler region. Using \eqref{eq:OPElimit}, one can also check that the OPE approximation is valid in this limit.

We now evaluate the entropy contribution from this island, expressing the result in Rindler coordinates. This results in an entropy that is time-independent,
\beq
S(R) = \frac c{6} + \frac{4\pi}{\kappa^2}(\Phi_0+\phi_0)+ \frac {c}3\log\left(\tfrac{\ell^2}{\epsilon_i\epsilon_r}\right)  ~.\label{eq:S(R)Before Pulse}
\eeq
We again rescaled the UV cutoffs as appropriate for Rindler coordinates, see \eqref{UVtrafo}.

One can also consider islands in the de Sitter region, in which case the entropy is given by
\beq
S(R) = \text{min~}\text{ext}_i\left[\frac{c}{6} + \frac{4\pi}{\kappa^2}\left(\Phi_0+\phi_0\frac{\ell^2+x^+_ix^-_i}{\ell^2-x^+_ix^-_i}\right) + \frac c3 \log\left[\frac{(x^+_i-x^+_r)(x^-_i-x^-_r)}{\Omega_{\rm dS}(x^+_i,x^-_i)\epsilon_i\epsilon_r}\right] \right]~.
\eeq
Extremizing the generalized entropy in the limit $|x^+_rx^-_r|\gg \ell^2$ and $c\ll \phi_{0}$, we find a quantum extremal surface at
\beq
x_i^\pm = \mp \frac{c\ell\kappa^2}{42\pi\phi_0}e^{\pm(t\mp\rho)/\ell}
~.\label{eq:islanddSBeforePulse}
\eeq
Note the crucial overall minus sign difference as compared to \eqref{eq:islandRindlerBeforePulse}.
In the late-time limit, the future horizon is reached from above. However, because $x^-_r\leq 0$ this island is of the minimax type and moves backward in time. As discussed in the introduction, this, therefore, leads to the pathological situation where basic entropy properties such as entanglement wedge nesting are violated \cite{Shaghoulian:2021cef}. We, therefore, omit these types of islands and only allow for a conventional maximin QES.

Thus, in this equilibrium situation, we find an entropy that is consistent with finite de Sitter entropy. However, because the islands are located in the Rindler region we don't seem to learn anything about de Sitter space. To probe de Sitter physics, we need to break the thermal equilibrium, which we will do next.

\subsubsection*{During the Pulse}
We will now consider islands that are located in the pulse region during $x^+_A\leq x^+_r \leq x^+_B$. This situation is similar to the setup in \cite{Aalsma:2021bit}, with the difference that our initial state at $x^+_r< x^+_A$ is in thermal equilibrium. We will impose that $x^+_B \ll x^+_{\rm end}$ (see \eqref{eq:BackLimit}) to ensure backreaction is under control. Because we work in the OPE limit, the total generalized entropy splits into two parts which we will denote by $S(R_L)$ and $S(R_R)$ corresponding to the entropy associated with the left and right observer respectively. In the OPE limit, the pulse can only affect the entropy of the right observer and we therefore only have to extremize this part of the entropy to study the effect of the pulse.

Focusing on an island in the de Sitter region, the entropy is given by
\beq\label{eq:Sgen(D_R) IP}
\bal
S_{\rm gen}(R_R) &= \frac{c}{12}\frac{\ell^2-x^+_Ax_i^-}{\ell^2-x_i^+x_i^-} + \frac{2\pi}{\kappa^2}\left(\Phi_0+\phi_0\frac{\ell^2+x_i^+x_i^-}{\ell^2-x^+_ix^-_i}\right) -\frac c{24}\frac{\ell^2+x_i^+x_i^-}{\ell^2-x_i^+x_i^-}\log\left(x_i^+/x^+_A\right) \\
&+ \frac c6 \log\left[\frac{\ell \log(x^+_i/x^+_r)(x^-_i-x^-_r)}{\Omega_{\rm dS}(x^+_i,x^-_i)\Omega_{\rm R}(x^+_r,x^-_r)\epsilon_i\epsilon_r}\right] ~.
\eal
\eeq
In order to find an approximate analytical expression for the quantum extremal surface, we will ignore left-moving modes in the entropy. This was also done in \cite{Aalsma:2021bit}, where it was confirmed this leads to a good approximation due to the absence of left-movers in the non-equilibrium state, and we find an island that is very similar,
\beq\label{eq:QES IP}
x^-_i = 0 ~, \quad x^+_i = -\frac{2\ell^2+x^+_Ax^-_r}{x^-_r}W_0[f(x^-_r)]^{-1} ~.
\eeq
Here $W_0(x)$ is the principal branch of the Lambert $W$ function with
\beq
f(x^-_r) = -\frac{2\ell^2+x^+_Ax^-_r}{x^+_Ax^-_r}\exp\left(-2-\frac{48\pi\phi_0}{c\kappa^2}\right) ~.
\eeq
We checked numerically that the ``true'' island lies at small but positive $x^-_i$. To get some intuition for the behavior of this island, we consider expanding the Lambert $W$ function. First, the smallest value of $x^+$ where the island can appear is given by
\beq \label{eq:SmallestIsland}
\lim_{x^-_r\to-\infty}x^+_i \simeq x^+_A e^{\frac{48\pi\phi_0}{c\kappa^2}} ~,
\eeq
where we made use of $\phi_0/c\gg 1$. Second, we can expand the Lambert $W$ function around infinity (corresponding to late times): $W_0[x] \simeq \log(x)$. Evaluating the entropy on this (doubly-)approximated island we find
\beq \label{eq:PulseEntropy}
S(R_R) \simeq \frac c{12} + \frac{2\pi}{\kappa^2}(\Phi_0+\phi_0) - \frac{c}{24\ell}(t_r-t_A) + \dots ~,
\eeq
where the dots denote unimportant constant contributions and the usual divergences.  We defined $t_A = \ell \log(x^+_A/\ell)$. This shows that, at least when the approximations are valid, the entropy $S(R_R)$ is equal to the de Sitter entropy at $t_r= t_A$ and then decreases linearly. We believe these approximations reveal the main qualitative features of the behavior of the entropy during this phase. In particular, it shows that we can decode information from behind the de Sitter horizon, due to the presence of an island in that region that probes the de Sitter geometry.

It is also possible to have an island located in the region before the pulse, but this island gives a subdominant (higher) entropy. There are no islands in the Rindler region.

\subsubsection*{After the Pulse}
After the pulse, a maximin island is present, only in the Rindler region, just as before the pulse. The generalized entropy is given by
\beq
\bal
S_{\rm gen}(R_R) &=\frac{c}{12} + \frac{2\pi}{\kappa^2}\left(\Phi_0+\phi_0(1-\ell^{-2}x^+x^-)\right) - \frac{c}{24}\log(x^+_B/x^+_A) \\
&+ \frac c6\log\left[\frac{(x^+_r-x^+_i)(x^-_i-x^-_r)}{\epsilon_i\epsilon_r}\right] ~.
\eal
\eeq
The only difference with the entropy before the pulse is the logarithmic factor in the dilaton solution, which is constant. Therefore, the location of the island is again given by
\beq\label{eq:QES AP}
\quad x_i^\pm = \mp\frac{c\kappa^2\ell^2}{12\pi\phi_0 x^-_r} ~,
\eeq
and evaluating the generalized entropy on this island in Rindler coordinates we obtain
\beq
S(R_R) = \frac c{12} + \frac{2\pi}{\kappa^2}(\Phi_0+\phi_0)   - \frac c{24}\log(x^+_B/x^+_A) + \frac {c}6\log\left(\tfrac{\ell^2}{\epsilon_r\epsilon_i}\right)~.
\eeq
Compared to the result before the pulse, we note that the de Sitter entropy has decreased by an amount
\beq
\Delta S =  \frac c{24}\log(x^+_B/x^+_A) ~,
\eeq
after the pulse. This matches the decrease of the entropy during the pulse phase as given by our estimate \eqref{eq:PulseEntropy}. This shows that by collecting radiation from the cosmological horizon in the Rindler region, we successfully decreased the entropy of de Sitter space, signaling information recovery.

We could also imagine the situation where the region $r$ is already located after the pulse ($x^+_r > x^+_B$), but there still is an island located in the pulse region. Such an island is indeed present, but at late times it increases and does not correspond to the dominant saddle.

\subsection{Controlled Information Recovery} \label{sec:InfoRecovery}
We now describe how the above results for the islands and generalized entropy leads to a controlled setup to recover information from behind the cosmological horizon. Before a pulse is performed that breaks the thermal equilibrium, the growing entropy associated with the trivial island saturates at
\beq \label{eq:tsat}
t_{\rm sat} \simeq 12\pi\ell(\Phi_0+\phi_0)/(c\kappa^2) ~.
\eeq
We can now consider two different situations, one where the pulse breaks the thermal equilibrium before this time scale, or after it. Let's imagine we want to decode a qubit of information at a time before the Page time. In this case, general arguments about the size of the quantum system we have access to \cite{Hayden:2007cs} suggest that after the qubit is thrown behind the cosmological horizon, it will take a time that scales linearly with the entropy before we can decode this qubit. However, if one first waits until an order one amount of the entropy has been emitted in terms of radiation, then throws in a qubit, decoding that qubit should be possible in a scrambling time. Our results are consistent with these arguments. We will consider both cases separately.

\subsubsection*{Pulse Before the Saturation Time}
\begin{figure}[t!]
\centering
\includegraphics[scale=.7]{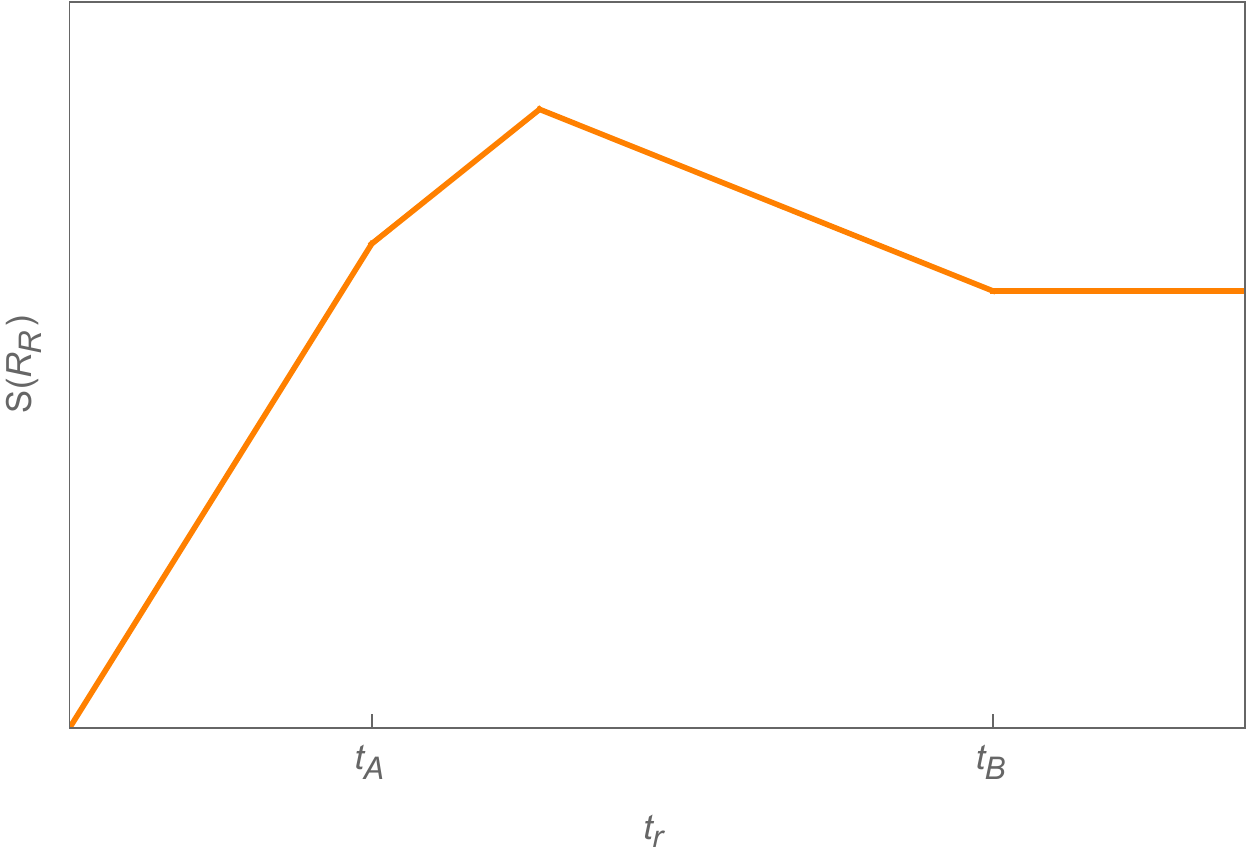}
\caption{Behavior of the entropy $S(R_R)$ in the scenario where a pulse is applied during the time $t_A\leq t_r\leq t_B$ before the saturation time is reached. The entropy is therefore always lower than the de Sitter entropy.}
\label{fig:PC1}
\end{figure}

Decoding becomes possible when the non-trivial island in the de Sitter region gives the dominant contribution to the entropy. However, from our estimate of the entropy in \eqref{eq:PulseEntropy}, we see that at the moment the pulse is applied the non-trivial de Sitter island gives an entropy equal to the de Sitter entropy and then decreases. Thus, when the pulse is applied before the saturation time is reached the dominant contribution during the pulse will be given by a trivial island. This corresponds to an increasing entropy that behaves as $S(R_R)= \frac{c}{12\ell}t$ (one-half the value of the trivial island before the pulse because there is no left-moving radiation). The entropy starts to decrease as $S(R_R)=\frac{c}{24}t$ when it becomes equal to the entropy of the non-trivial island.\footnote{One could imagine the situation where the pulse ends before the entropy given by the trivial island during the pulse intersects the decreasing entropy. In this case, the entropy saturates at a constant value after the pulse, but no information recovery is possible.} This happens only after a time that scales linearly with the entropy. We plot the entropy $S(R_R)$ in this scenario in Figure \ref{fig:PC1}.

\subsubsection*{Pulse After Saturation Time}
\begin{figure}[t!]
\centering
\includegraphics[scale=.7]{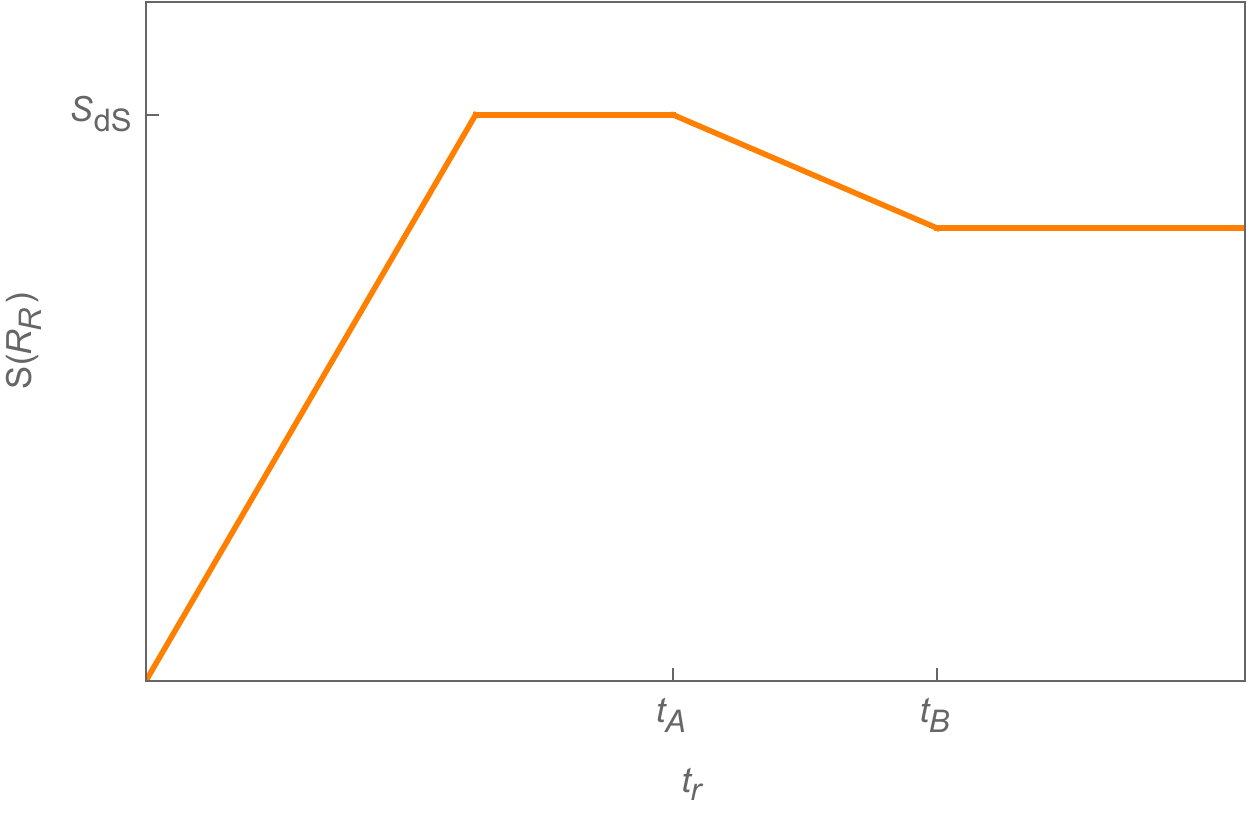}
\caption{Behavior of the entropy $S(R_R)$ in the scenario where a pulse is applied during the time $t_A\leq t_r\leq t_B$ after the saturation time is reached. The highest plateau corresponds to the de Sitter entropy $S_{\rm dS}=\frac{2\pi}{\kappa^2}(\Phi_0+\Phi)_{x^-=0}$.}
\label{fig:PC2}
\end{figure}

Now consider the situation that we first wait until the saturation time before the pulse. In this case, the non-equilibrium island can give the dominant contribution when the pulse is applied and information recovery becomes possible much more quickly. In this case, to recover some information we only have to wait for the difference between the location of the island and the location of the asymptotic observer. We measure this by
\beq
\Delta t= \ell \log\left(x^+_r/x^+_i\right) ~.
\eeq
Up to logarithmic corrections, the evaluation of \eqref{eq:QES IP} at late times, where we approximate $W_0[x] \simeq \log(x)$, leads to
\beq
\Delta t \simeq  \ell\log(t_r/\ell) + 2\rho_r ~.
\eeq
We see that the time difference between information reaching the island and its recovery scales logarithmically with time and contains a radial piece $2\rho_r/\ell$ that is associated with the time it takes to travel to the anchor curve \cite{Gautason:2020tmk}. Because information recovery becomes possible around the saturation time \eqref{eq:tsat}, we see that the first part of $\Delta t$ evaluates to the scrambling time:
\beq
\Delta t \simeq \ell\log(S_{\rm dS}/c) + 2\rho_r ~,
\eeq
where we again dropped a small logarithmic correction. The behavior of the entropy in this scenario is plotted in Figure \ref{fig:PC2}.

\subsubsection*{Backreaction Effects}\label{sec:Backreaction effects}
To argue that the above scenarios indeed correspond to controlled information recovery requires backreaction effects to be small. For instance, while \cite{Aalsma:2021bit} showed that a non-equilibrium state similar to the one used here allows for information recovery, backreaction effects eventually become large and the formation of a singularity that destroys the static patch is unavoidable. In contrast, the setup we studied in this paper restores a thermal equilibrium after $x^+>x^+_B$, effectively cutting off catastrophic backreaction. 

Nonetheless, during the pulse a singularity can possibly form at ${\cal I}^+$ as the dilaton behaves as
\beq
\lim_{x^+x^-\to\ell^2} \Phi =(\infty) \times \text{sgn}\left(c(x^+-x^+_A)\kappa^2+48\pi\phi_0 x^+-c\kappa^2x^+\log(x^+/x^+_A)\right) ~.
\eeq
Thus, when the sign is negative a singularity will form at ${\cal I}^+$. For large $x^+/\ell$ and $\phi_0/c$, this will occur at
\beq
x^+\geq x^+_{\rm sing} = x^+_A e^{\frac{48\pi\phi_0}{c\kappa^2}} ~.
\eeq
We note that this coincides with the smallest possible value of $x^+$ for the island (see \eqref{eq:SmallestIsland}). This implies that when information recovery becomes possible, a singularity is guaranteed to form at ${\cal I}^+$. Nonetheless, this singularity is confined to the region at ${\cal I}^+$ where $x^+_{\rm sing}\leq x^+\leq x^+_B$. Therefore, an observer performing a decoding operation is able to survive this experiment. The amount to which the singularity extends along ${\cal I}^+$ is directly proportional to the energy and entropy collected. Thus, by controlling the duration of the pulse, the amount of backreaction is controllable.

\subsection{Remarks on the No-Cloning Theorem} \label{sec:No-Cloning}
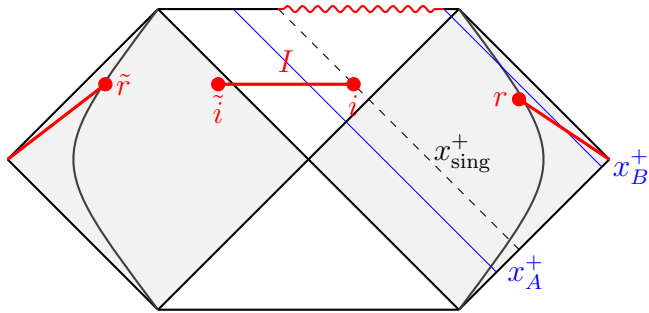
\begin{figure}[t!]
\centering
\begin{tikzpicture}

\fill [gray!10,rotate around={-45:(4,0)}] (4,0) rectangle (6.8,2.8);
\fill [gray!10,rotate around={135:(4,0)}] (4,0) rectangle (6.8,2.8);

\draw[thick] (0,0) -- (2,2);
\draw[thick] (0,0) -- (2,-2);
\draw[thick] (2,2) -- (6,-2);
\draw[thick] (2,-2) -- (6,2);
\draw[thick] (2,2) -- (3.6,2);
\draw[thick] (5.8,2) -- (6,2);
\draw[thick] (2,-2) -- (6,-2);
\draw[thick] (6,2) -- (8,0);
\draw[thick] (6,-2) -- (8,0);

\draw[thick,darkgray] (6,-2) .. controls (7.5,0) .. (6,2);
\draw[thick,darkgray] (2,-2) .. controls (0.5,0) .. (2,2);

\draw[blue] (3,2) -- (6.5,-1.5) node[right] {$x^+_A$};
\draw[blue] (5.8,2) -- (7.9,-.1) node[right] {$x^+_B$};

\draw[dashed] (3.6,2) -- (6.8,-1.2) node[right,pos=.6]{$x^+_{\rm sing}$};

\draw[red,very thick] (6.8,.8) -- (8,0);
\filldraw[red] (6.8,.8) circle (2.5pt) node[left] {$r$};
\draw[red,very thick] (0,0) -- (1.3,1);
\filldraw[red] (1.3,1) circle (2.5pt) node[right] {$\tilde r$};

\draw[red, very thick] (2.8,1) -- (4.6,1) node[pos=.5,above]{$I$};
\filldraw[red] (2.8,1) circle (2.5pt) node[below] {$\tilde i$};
\filldraw[red] (4.6,1) circle (2.5pt) node[below] {$i$};

\tikzset{decoration={snake,amplitude=.4mm,segment length=2mm,
                       post length=.6mm,pre length=.6mm}};
\draw[red,thick,decorate] (3.6,2) -- (5.8,2);

\end{tikzpicture}
\caption{During the pulse ($x^+_A\leq x^+\leq x^+_B$), the location $x^+_i$ of the de Sitter island that gives a decreasing entropy is always bigger than or equal to the location $x^+_{\rm sing}$ where a singularity forms at ${\cal I}^+$.}
\label{fig:EW}
\end{figure}

One major difference between information recovery in the cosmological model we studied in this paper and an evaporating black hole is the location of the singularity behind the horizon. Whereas a geodesic that crosses the horizon of a black hole is guaranteed to hit a singularity, the singularity in our model at ${\cal I}^+$ is confined to lie at $x^+_{\rm sing}\leq x^+\leq x^+_B$, see Figure \ref{fig:EW}.
This potentially leads to a paradox concerning a violation of the no-cloning theorem \cite{Dieks:1982dj}. In a well-known paper \cite{Hayden:2007cs}, predating the current understanding of entanglement islands, Hayden and Preskill considered if decoding information from Hawking radiation can lead to an observable violation of the no-cloning theorem. Their idea was that when a message, such as qubit, is thrown inside a black hole and afterward decoded from the Hawking radiation it is being copied. If an observer, after decoding, jumps inside the black hole they might observe the same qubit twice, in violation of no-cloning. However, there is a danger that either the original copy or the observer hits the singularity before this violation can be observed. Avoiding this situation requires a trans-Planckian fine-tuning, and \cite{Hayden:2007cs} therefore concluded that the no-cloning theorem is safe. In contrast, in our cosmological model there is no singularity behind the horizon at $x^+>x^+_B$ and no-cloning might be in jeopardy again. We will give two complementary points of view to argue that this is not the case.

First, setting up a thought experiment that has the potential to violate no-cloning in some sense assumes that the region behind the horizon is independent of the radiation that is being emitted. This seems to be a necessary ingredient when one argues that a message is being copied when it is being decoded from Hawking radiation. However, one of the main lessons of entanglement islands is that this premise is not entirely true. The fact that the entanglement wedge of Hawking radiation can include a disconnected region behind the horizon shows that the two cannot be thought of as completely independent. In that sense, the reasoning that leads one to conclude that a message is being copied is, given our current understanding of islands, flawed. 

Second, decoding a message is possible when the entropy decreases which, in our setup, happens when the non-trivial island in the de Sitter region gives the dominant contribution to entropy. Decoding a qubit is therefore only possible during the pulse or, said differently, when the entanglement wedge of the radiation includes the de Sitter island. However, as we remarked before, a singularity is present during this phase at ${\cal I}^+$ and the arguments of \cite{Hayden:2007cs} against observable violations of no-cloning apply, see Figure \ref{fig:EW}. After the qubit reaches $x^+>x^+_B$, the Rindler island gives the dominant contribution again, leading to a saturated entropy but no decoding. Therefore, it has ``escaped'' decoding and is not cloned.

\section{Discussion} \label{sec:discussion}
The aim of this paper was to understand information recovery in de Sitter space by an observer that collects radiation emanating from the cosmological horizon. In particular, we made use of a non-equilibrium quantum state, previously studied in \cite{Aalsma:2021bit}, to show that a conventional maximin QES can appear in de Sitter space. The main advantage of the setup studied in this paper as compared to \cite{Aalsma:2021bit} is that we dynamically connected this non-equilibrium state to the Bunch-Davies vacuum. This allowed us to break the thermal equilibrium only briefly, effectively cutting off the large and catastrophic backreaction present when the non-equilibrium state is turned on for a prolonged period.

The static patch of de Sitter space has no asymptotic weakly-gravitating regions where the island formula can be naturally applied, so we introduced those in a somewhat involved manner. We considered two different two-dimensional gravity solutions corresponding to de Sitter and Rindler space and introduced a domain wall that interpolated between these two models. By solving the two-dimensional analog of the Israel junction conditions, we obtained a geometry in which the static patches of de Sitter space are effectively replaced by Rindler wedges. Nonetheless, this setup allows us to probe the de Sitter region, as we demonstrated by the presence of an island in the de Sitter region. We briefly commented upon the higher-dimensional interpretation of this model, but leave a more detailed study to future work.

Our setup enables an observer, far away from the cosmological horizon where gravity is weak, to briefly break the thermal equilibrium and collect radiation that is emitted from the cosmological horizon. By computing its entropy using the island formula we show that information from beyond the cosmological horizon can be recovered, rather analogously as for an evaporating asymptotically flat black hole. Importantly, in this model the thermal equilibrium is only broken for a finite time, an essential difference with \cite{Aalsma:2021bit}, which keeps backreaction under control. This shows that information can be recovered from behind the cosmological horizon in a controlled manner, albeit at the cost of introducing Rindler wedges.

Although our model has strong backreaction if we keep the pulse turned on all the way to $x^{+}=x^{+}_{\text{end}}$, this endpoint is alike the final stage of an evaporating black hole in the RST model \cite{Russo:1992ax}. There, it was shown that once the black hole has entirely evaporated a shockwave with a negative energy of order $c$ is released, which was dubbed a ``thunderpop'' in contrast to the stronger ``thunderbolt''. This is only a relatively small amount of energy that does not modify the geometry in a dramatic way. It would be interesting to understand if something similar happens in the current model.

We conclude with some future directions. In the current work, we mostly focused on the entropy associated with the right region ${\cal C}_R$, but one can also consider information recovery from the left region. In particular, a qubit of information can be decoded either in the left or right region and it is a priori not clear which observer has access to this information. One peculiar feature of both observers applying a pulse is that the region where both pulses intersect is in the static vacuum, which does not contain any radiation. It could be useful to study whether or not information can be recovered from such a region and which observer would claim it. A similar situation was found in \cite{Almheiri:2019yqk}.

Furthermore, our point of view has been entirely from a Rindler observer. It would be interesting to work out the implications of our results for an observer that only has access to ${\cal I}^+$. One can think for example of an observer whose past lightcone contains a period of inflation and this scenario, therefore, more naturally connects to inflationary physics. Previous work suggests the presence of timelike separated islands \cite{Chen:2020tes} in such a context whose significance is not entirely clear, although there is recent work in this direction \cite{Doi:2022iyj}. 

Finally, one hope is that two-dimensional cosmological models, like the one studied in this paper, will be useful to better understand the holographic principle in cosmology. In particular, it is not completely understood how holographic properties like entanglement wedge reconstruction, quantum error correction and probes like quantum chaos \cite{Anninos:2018svg,Aalsma:2020aib,Blommaert:2020tht} and complexity \cite{Chapman:2021eyy,Jorstad:2022mls} manifest themselves in de Sitter space. We believe two-dimensional gravity will play a key role in answering these questions.

\subsection*{Acknowledgements}
We are grateful to the organizers of the ``Back to the Swamp'' workshop, where part of this work was presented, for a stimulating environment. LA and SEAG thank the University of Amsterdam where the final stages of this work were performed. We'd also like to thank Ben Freivogel, Friðrik Gautason, Juan Hernandez, Mikhail Khramtsov, Maria Knysh, Jake McNamara, Miguel Montero, Rahul Poddar, Tom Rudelius and Jan Pieter van der Schaar for useful discussions. LA is supported by the Heising-Simons Foundation under the “Observational Signatures of Quantum Gravity” collaboration grant 2021-2818. SEAG is supported by the KU Leuven C1 grant ZKD1118 C16/16/005. WS is supported by the Icelandic Research Fund via the Grant of Excellence titled “Quantum Fields and Quantum Geometry”.

\bibliographystyle{JHEP}
\bibliography{refs}

\providecommand{\href}[2]{#2}\begingroup\raggedright\begin{thebibliography}{10}

\bibitem{Engelhardt:2014gca}
N.~Engelhardt and A.C.~Wall, \emph{{Quantum Extremal Surfaces: Holographic
  Entanglement Entropy beyond the Classical Regime}},
  \href{https://doi.org/10.1007/JHEP01(2015)073}{\emph{JHEP} {\bfseries 01}
  (2015) 073} [\href{https://arxiv.org/abs/1408.3203}{{\ttfamily 1408.3203}}].

\bibitem{Penington:2019npb}
G.~Penington, \emph{{Entanglement Wedge Reconstruction and the Information
  Paradox}}, \href{https://doi.org/10.1007/JHEP09(2020)002}{\emph{JHEP}
  {\bfseries 09} (2020) 002}
  [\href{https://arxiv.org/abs/1905.08255}{{\ttfamily 1905.08255}}].

\bibitem{Almheiri:2019psf}
A.~Almheiri, N.~Engelhardt, D.~Marolf and H.~Maxfield, \emph{{The entropy of
  bulk quantum fields and the entanglement wedge of an evaporating black
  hole}}, \href{https://doi.org/10.1007/JHEP12(2019)063}{\emph{JHEP} {\bfseries
  12} (2019) 063} [\href{https://arxiv.org/abs/1905.08762}{{\ttfamily
  1905.08762}}].

\bibitem{Page:1993wv}
D.N.~Page, \emph{{Information in black hole radiation}},
  \href{https://doi.org/10.1103/PhysRevLett.71.3743}{\emph{Phys. Rev. Lett.}
  {\bfseries 71} (1993) 3743}
  [\href{https://arxiv.org/abs/hep-th/9306083}{{\ttfamily hep-th/9306083}}].

\bibitem{Almheiri:2019psy}
A.~Almheiri, R.~Mahajan and J.E.~Santos, \emph{{Entanglement islands in higher
  dimensions}},
  \href{https://doi.org/10.21468/SciPostPhys.9.1.001}{\emph{SciPost Phys.}
  {\bfseries 9} (2020) 001} [\href{https://arxiv.org/abs/1911.09666}{{\ttfamily
  1911.09666}}].

\bibitem{Laddha:2020kvp}
A.~Laddha, S.G.~Prabhu, S.~Raju and P.~Shrivastava, \emph{{The Holographic
  Nature of Null Infinity}},
  \href{https://doi.org/10.21468/SciPostPhys.10.2.041}{\emph{SciPost Phys.}
  {\bfseries 10} (2021) 041}
  [\href{https://arxiv.org/abs/2002.02448}{{\ttfamily 2002.02448}}].

\bibitem{Geng:2020qvw}
H.~Geng and A.~Karch, \emph{{Massive islands}},
  \href{https://doi.org/10.1007/JHEP09(2020)121}{\emph{JHEP} {\bfseries 09}
  (2020) 121} [\href{https://arxiv.org/abs/2006.02438}{{\ttfamily
  2006.02438}}].

\bibitem{Chowdhury:2020hse}
C.~Chowdhury, O.~Papadoulaki and S.~Raju, \emph{{A physical protocol for
  observers near the boundary to obtain bulk information in quantum gravity}},
  \href{https://doi.org/10.21468/SciPostPhys.10.5.106}{\emph{SciPost Phys.}
  {\bfseries 10} (2021) 106}
  [\href{https://arxiv.org/abs/2008.01740}{{\ttfamily 2008.01740}}].

\bibitem{Geng:2020fxl}
H.~Geng, A.~Karch, C.~Perez-Pardavila, S.~Raju, L.~Randall, M.~Riojas et~al.,
  \emph{{Information Transfer with a Gravitating Bath}},
  \href{https://doi.org/10.21468/SciPostPhys.10.5.103}{\emph{SciPost Phys.}
  {\bfseries 10} (2021) 103}
  [\href{https://arxiv.org/abs/2012.04671}{{\ttfamily 2012.04671}}].

\bibitem{Geng:2021hlu}
H.~Geng, A.~Karch, C.~Perez-Pardavila, S.~Raju, L.~Randall, M.~Riojas et~al.,
  \emph{{Inconsistency of islands in theories with long-range gravity}},
  \href{https://doi.org/10.1007/JHEP01(2022)182}{\emph{JHEP} {\bfseries 01}
  (2022) 182} [\href{https://arxiv.org/abs/2107.03390}{{\ttfamily
  2107.03390}}].

\bibitem{Chowdhury:2021nxw}
C.~Chowdhury, V.~Godet, O.~Papadoulaki and S.~Raju, \emph{{Holography from the
  Wheeler-DeWitt equation}},
  \href{https://doi.org/10.1007/JHEP03(2022)019}{\emph{JHEP} {\bfseries 03}
  (2022) 019} [\href{https://arxiv.org/abs/2107.14802}{{\ttfamily
  2107.14802}}].

\bibitem{Raju:2021lwh}
S.~Raju, \emph{{Failure of the split property in gravity and the information
  paradox}}, \href{https://doi.org/10.1088/1361-6382/ac482b}{\emph{Class.
  Quant. Grav.} {\bfseries 39} (2022) 064002}
  [\href{https://arxiv.org/abs/2110.05470}{{\ttfamily 2110.05470}}].

\bibitem{Bousso:2022hlz}
R.~Bousso and G.~Penington, \emph{{Entanglement Wedges for Gravitating
  Regions}},  \href{https://arxiv.org/abs/2208.04993}{{\ttfamily 2208.04993}}.

\bibitem{Aalsma:2021bit}
L.~Aalsma and W.~Sybesma, \emph{{The Price of Curiosity: Information Recovery
  in de Sitter Space}},
  \href{https://doi.org/10.1007/JHEP05(2021)291}{\emph{JHEP} {\bfseries 05}
  (2021) 291} [\href{https://arxiv.org/abs/2104.00006}{{\ttfamily
  2104.00006}}].

\bibitem{Kames-King:2021etp}
J.~Kames-King, E.M.H.~Verheijden and E.P.~Verlinde, \emph{{No Page curves for
  the de Sitter horizon}},
  \href{https://doi.org/10.1007/JHEP03(2022)040}{\emph{JHEP} {\bfseries 03}
  (2022) 040} [\href{https://arxiv.org/abs/2108.09318}{{\ttfamily
  2108.09318}}].

\bibitem{Gautason:2020tmk}
F.F.~Gautason, L.~Schneiderbauer, W.~Sybesma and L.~Thorlacius, \emph{{Page
  Curve for an Evaporating Black Hole}},
  \href{https://doi.org/10.1007/JHEP05(2020)091}{\emph{JHEP} {\bfseries 05}
  (2020) 091} [\href{https://arxiv.org/abs/2004.00598}{{\ttfamily
  2004.00598}}].

\bibitem{Hartman:2020swn}
T.~Hartman, E.~Shaghoulian and A.~Strominger, \emph{{Islands in Asymptotically
  Flat 2D Gravity}}, \href{https://doi.org/10.1007/JHEP07(2020)022}{\emph{JHEP}
  {\bfseries 07} (2020) 022}
  [\href{https://arxiv.org/abs/2004.13857}{{\ttfamily 2004.13857}}].

\bibitem{Balasubramanian:2020coy}
V.~Balasubramanian, A.~Kar and T.~Ugajin, \emph{{Entanglement between two
  disjoint universes}},
  \href{https://doi.org/10.1007/JHEP02(2021)136}{\emph{JHEP} {\bfseries 02}
  (2021) 136} [\href{https://arxiv.org/abs/2008.05274}{{\ttfamily
  2008.05274}}].

\bibitem{Miyata:2021ncm}
A.~Miyata and T.~Ugajin, \emph{{Evaporation of black holes in flat space
  entangled with an auxiliary universe}},
  \href{https://doi.org/10.1093/ptep/ptab163}{\emph{PTEP} {\bfseries 2022}
  (2022) 013B13} [\href{https://arxiv.org/abs/2104.00183}{{\ttfamily
  2104.00183}}].

\bibitem{Balasubramanian:2021wgd}
V.~Balasubramanian, A.~Kar and T.~Ugajin, \emph{{Entanglement between two
  gravitating universes}},  \href{https://arxiv.org/abs/2104.13383}{{\ttfamily
  2104.13383}}.

\bibitem{Chandrasekaran:2021tkb}
V.~Chandrasekaran, T.~Faulkner and A.~Levine, \emph{{Scattering strings off
  quantum extremal surfaces}},
  \href{https://doi.org/10.1007/JHEP08(2022)143}{\emph{JHEP} {\bfseries 08}
  (2022) 143} [\href{https://arxiv.org/abs/2108.01093}{{\ttfamily
  2108.01093}}].

\bibitem{Miyata:2021qsm}
A.~Miyata and T.~Ugajin, \emph{{Entanglement between two evaporating black
  holes}},  \href{https://arxiv.org/abs/2111.11688}{{\ttfamily 2111.11688}}.

\bibitem{DeVuyst:2022bua}
J.~De~Vuyst and T.G.~Mertens, \emph{{Operational islands and black hole
  dissipation in JT gravity}},
  \href{https://arxiv.org/abs/2207.03351}{{\ttfamily 2207.03351}}.

\bibitem{Yu:2022xlh}
M.-H.~Yu and X.-H.~Ge, \emph{{Entanglement Islands in Generalized
  Two-dimensional Dilaton Black Holes}},
  \href{https://arxiv.org/abs/2208.01943}{{\ttfamily 2208.01943}}.

\bibitem{Murdia:2022giv}
C.~Murdia, Y.~Nomura and K.~Ritchie, \emph{{Black Hole and de Sitter
  Microstructures from a Semiclassical Perspective}},
  \href{https://arxiv.org/abs/2207.01625}{{\ttfamily 2207.01625}}.

\bibitem{Chen:2020tes}
Y.~Chen, V.~Gorbenko and J.~Maldacena, \emph{{Bra-ket wormholes in
  gravitationally prepared states}},
  \href{https://doi.org/10.1007/JHEP02(2021)009}{\emph{JHEP} {\bfseries 02}
  (2021) 009} [\href{https://arxiv.org/abs/2007.16091}{{\ttfamily
  2007.16091}}].

\bibitem{Hartman:2020khs}
T.~Hartman, Y.~Jiang and E.~Shaghoulian, \emph{{Islands in cosmology}},
  \href{https://doi.org/10.1007/JHEP11(2020)111}{\emph{JHEP} {\bfseries 11}
  (2020) 111} [\href{https://arxiv.org/abs/2008.01022}{{\ttfamily
  2008.01022}}].

\bibitem{Sybesma:2020fxg}
W.~Sybesma, \emph{{Pure de Sitter space and the island moving back in time}},
  \href{https://doi.org/10.1088/1361-6382/abff9a}{\emph{Class. Quant. Grav.}
  {\bfseries 38} (2021) 145012}
  [\href{https://arxiv.org/abs/2008.07994}{{\ttfamily 2008.07994}}].

\bibitem{Balasubramanian:2020xqf}
V.~Balasubramanian, A.~Kar and T.~Ugajin, \emph{{Islands in de Sitter space}},
  \href{https://doi.org/10.1007/JHEP02(2021)072}{\emph{JHEP} {\bfseries 02}
  (2021) 072} [\href{https://arxiv.org/abs/2008.05275}{{\ttfamily
  2008.05275}}].

\bibitem{Geng:2021wcq}
H.~Geng, Y.~Nomura and H.-Y.~Sun, \emph{{Information paradox and its resolution
  in de Sitter holography}},
  \href{https://doi.org/10.1103/PhysRevD.103.126004}{\emph{Phys. Rev. D}
  {\bfseries 103} (2021) 126004}
  [\href{https://arxiv.org/abs/2103.07477}{{\ttfamily 2103.07477}}].

\bibitem{Aguilar-Gutierrez:2021bns}
S.E.~Aguilar-Gutierrez, A.~Chatwin-Davies, T.~Hertog, N.~Pinzani-Fokeeva and
  B.~Robinson, \emph{{Islands in Multiverse Models}},
  \href{https://doi.org/10.1007/JHEP11(2021)212}{\emph{JHEP} {\bfseries 11}
  (2021) 212} [\href{https://arxiv.org/abs/2108.01278}{{\ttfamily
  2108.01278}}].

\bibitem{Langhoff:2021uct}
K.~Langhoff, C.~Murdia and Y.~Nomura, \emph{{Multiverse in an inverted
  island}}, \href{https://doi.org/10.1103/PhysRevD.104.086007}{\emph{Phys. Rev.
  D} {\bfseries 104} (2021) 086007}
  [\href{https://arxiv.org/abs/2106.05271}{{\ttfamily 2106.05271}}].

\bibitem{Goswami:2021ksw}
K.~Goswami, K.~Narayan and H.K.~Saini, \emph{{Cosmologies, singularities and
  quantum extremal surfaces}},
  \href{https://doi.org/10.1007/JHEP03(2022)201}{\emph{JHEP} {\bfseries 03}
  (2022) 201} [\href{https://arxiv.org/abs/2111.14906}{{\ttfamily
  2111.14906}}].

\bibitem{Bousso:2022gth}
R.~Bousso and E.~Wildenhain, \emph{{Islands in closed and open universes}},
  \href{https://doi.org/10.1103/PhysRevD.105.086012}{\emph{Phys. Rev. D}
  {\bfseries 105} (2022) 086012}
  [\href{https://arxiv.org/abs/2202.05278}{{\ttfamily 2202.05278}}].

\bibitem{Espindola:2022fqb}
R.~Esp\'\i{}ndola, B.~Najian and D.~Nikolakopoulou, \emph{{Islands in FRW
  Cosmologies}},  \href{https://arxiv.org/abs/2203.04433}{{\ttfamily
  2203.04433}}.

\bibitem{Svesko:2022txo}
A.~Svesko, E.~Verheijden, E.P.~Verlinde and M.R.~Visser, \emph{{Quasi-local
  energy and microcanonical entropy in two-dimensional nearly de Sitter
  gravity}}, \href{https://doi.org/10.1007/JHEP08(2022)075}{\emph{JHEP}
  {\bfseries 08} (2022) 075}
  [\href{https://arxiv.org/abs/2203.00700}{{\ttfamily 2203.00700}}].

\bibitem{Levine:2022wos}
A.~Levine and E.~Shaghoulian, \emph{{Encoding beyond cosmological horizons in
  de Sitter JT gravity}},  \href{https://arxiv.org/abs/2204.08503}{{\ttfamily
  2204.08503}}.

\bibitem{Azarnia:2022kmp}
S.~Azarnia and R.~Fareghbal, \emph{{Islands in Kerr\textendash{}de Sitter
  spacetime and their flat limit}},
  \href{https://doi.org/10.1103/PhysRevD.106.026012}{\emph{Phys. Rev. D}
  {\bfseries 106} (2022) 026012}
  [\href{https://arxiv.org/abs/2204.08488}{{\ttfamily 2204.08488}}].

\bibitem{Yadav:2022jib}
G.~Yadav and N.~Joshi, \emph{{Cosmological and black hole Islands in
  multi-event horizon spacetimes}},
  \href{https://arxiv.org/abs/2210.00331}{{\ttfamily 2210.00331}}.

\bibitem{Goswami:2022ylc}
K.~Goswami and K.~Narayan, \emph{{Small Schwarzschild de Sitter black holes,
  quantum extremal surfaces and islands}},
  \href{https://arxiv.org/abs/2207.10724}{{\ttfamily 2207.10724}}.

\bibitem{Almheiri:2019hni}
A.~Almheiri, R.~Mahajan, J.~Maldacena and Y.~Zhao, \emph{{The Page curve of
  Hawking radiation from semiclassical geometry}},
  \href{https://doi.org/10.1007/JHEP03(2020)149}{\emph{JHEP} {\bfseries 03}
  (2020) 149} [\href{https://arxiv.org/abs/1908.10996}{{\ttfamily
  1908.10996}}].

\bibitem{Shaghoulian:2021cef}
E.~Shaghoulian, \emph{{The central dogma and cosmological horizons}},
  \href{https://doi.org/10.1007/JHEP01(2022)132}{\emph{JHEP} {\bfseries 01}
  (2022) 132} [\href{https://arxiv.org/abs/2110.13210}{{\ttfamily
  2110.13210}}].

\bibitem{Mahajan21}
R.~Mahajan and E.~Shaghoulian, ``{Islands and closed universes}.''
  \url{https://raghumahajan.files.wordpress.com/2021/06/bu-2.pdf}, 2021.

\bibitem{Doi:2022iyj}
K.~Doi, J.~Harper, A.~Mollabashi, T.~Takayanagi and Y.~Taki, \emph{{Pseudo
  Entropy in dS/CFT and Time-like Entanglement Entropy}},
  \href{https://arxiv.org/abs/2210.09457}{{\ttfamily 2210.09457}}.

\bibitem{Aalsma:2021kle}
L.~Aalsma, A.~Cole, E.~Morvan, J.P.~van~der Schaar and G.~Shiu, \emph{{Shocks
  and information exchange in de Sitter space}},
  \href{https://doi.org/10.1007/JHEP10(2021)104}{\emph{JHEP} {\bfseries 10}
  (2021) 104} [\href{https://arxiv.org/abs/2105.12737}{{\ttfamily
  2105.12737}}].

\bibitem{Banihashemi:2022htw}
B.~Banihashemi, T.~Jacobson, A.~Svesko and M.~Visser, \emph{{The minus sign in
  the first law of de Sitter horizons}},
  \href{https://arxiv.org/abs/2208.11706}{{\ttfamily 2208.11706}}.

\bibitem{Teitelboim:1983ux}
C.~Teitelboim, \emph{{Gravitation and Hamiltonian Structure in Two Space-Time
  Dimensions}}, \href{https://doi.org/10.1016/0370-2693(83)90012-6}{\emph{Phys.
  Lett. B} {\bfseries 126} (1983) 41}.

\bibitem{Jackiw:1984je}
R.~Jackiw, \emph{{Lower Dimensional Gravity}},
  \href{https://doi.org/10.1016/0550-3213(85)90448-1}{\emph{Nucl. Phys. B}
  {\bfseries 252} (1985) 343}.

\bibitem{Callan:1992rs}
C.G.~Callan, Jr., S.B.~Giddings, J.A.~Harvey and A.~Strominger,
  \emph{{Evanescent black holes}},
  \href{https://doi.org/10.1103/PhysRevD.45.R1005}{\emph{Phys. Rev. D}
  {\bfseries 45} (1992) R1005}
  [\href{https://arxiv.org/abs/hep-th/9111056}{{\ttfamily hep-th/9111056}}].

\bibitem{Christensen:1977jc}
S.M.~Christensen and S.A.~Fulling, \emph{{Trace Anomalies and the Hawking
  Effect}}, \href{https://doi.org/10.1103/PhysRevD.15.2088}{\emph{Phys. Rev. D}
  {\bfseries 15} (1977) 2088}.

\bibitem{Pedraza:2021cvx}
J.F.~Pedraza, A.~Svesko, W.~Sybesma and M.R.~Visser, \emph{{Semi-classical
  thermodynamics of quantum extremal surfaces in Jackiw-Teitelboim gravity}},
  \href{https://doi.org/10.1007/JHEP12(2021)134}{\emph{JHEP} {\bfseries 12}
  (2021) 134} [\href{https://arxiv.org/abs/2107.10358}{{\ttfamily
  2107.10358}}].

\bibitem{Pedraza:2021ssc}
J.F.~Pedraza, A.~Svesko, W.~Sybesma and M.R.~Visser, \emph{{Microcanonical
  action and the entropy of Hawking radiation}},
  \href{https://doi.org/10.1103/PhysRevD.105.126010}{\emph{Phys. Rev. D}
  {\bfseries 105} (2022) 126010}
  [\href{https://arxiv.org/abs/2111.06912}{{\ttfamily 2111.06912}}].

\bibitem{Godet:2021cdl}
V.~Godet and C.~Marteau, \emph{{From black holes to baby universes in CGHS
  gravity}}, \href{https://doi.org/10.1007/JHEP07(2021)138}{\emph{JHEP}
  {\bfseries 07} (2021) 138}
  [\href{https://arxiv.org/abs/2103.13422}{{\ttfamily 2103.13422}}].

\bibitem{Engelhardt:2022qts}
N.~Engelhardt and r.~Folkestad, \emph{{Canonical purification of evaporating
  black holes}}, \href{https://doi.org/10.1103/PhysRevD.105.086010}{\emph{Phys.
  Rev. D} {\bfseries 105} (2022) 086010}
  [\href{https://arxiv.org/abs/2201.08395}{{\ttfamily 2201.08395}}].

\bibitem{Mirbabayi:2020grb}
M.~Mirbabayi, \emph{{Uptunneling to de Sitter}},
  \href{https://doi.org/10.1007/JHEP09(2020)070}{\emph{JHEP} {\bfseries 09}
  (2020) 070} [\href{https://arxiv.org/abs/2003.05460}{{\ttfamily
  2003.05460}}].

\bibitem{Maldacena:2019cbz}
J.~Maldacena, G.J.~Turiaci and Z.~Yang, \emph{{Two dimensional Nearly de Sitter
  gravity}}, \href{https://doi.org/10.1007/JHEP01(2021)139}{\emph{JHEP}
  {\bfseries 01} (2021) 139}
  [\href{https://arxiv.org/abs/1904.01911}{{\ttfamily 1904.01911}}].

\bibitem{Giddings:1992kn}
S.B.~Giddings and A.~Strominger, \emph{{Dynamics of extremal black holes}},
  \href{https://doi.org/10.1103/PhysRevD.46.627}{\emph{Phys. Rev. D} {\bfseries
  46} (1992) 627} [\href{https://arxiv.org/abs/hep-th/9202004}{{\ttfamily
  hep-th/9202004}}].

\bibitem{Mazur:2001fv}
P.O.~Mazur and E.~Mottola, \emph{{Gravitational condensate stars: An
  alternative to black holes}},
  \href{https://arxiv.org/abs/gr-qc/0109035}{{\ttfamily gr-qc/0109035}}.

\bibitem{Aalsma:2019rpt}
L.~Aalsma, M.~Parikh and J.P.~Van Der~Schaar, \emph{{Back(reaction) to the
  Future in the Unruh-de Sitter State}},
  \href{https://doi.org/10.1007/JHEP11(2019)136}{\emph{JHEP} {\bfseries 11}
  (2019) 136} [\href{https://arxiv.org/abs/1905.02714}{{\ttfamily
  1905.02714}}].

\bibitem{Fiola:1994ir}
T.M.~Fiola, J.~Preskill, A.~Strominger and S.P.~Trivedi, \emph{{Black hole
  thermodynamics and information loss in two-dimensions}},
  \href{https://doi.org/10.1103/PhysRevD.50.3987}{\emph{Phys. Rev. D}
  {\bfseries 50} (1994) 3987}
  [\href{https://arxiv.org/abs/hep-th/9403137}{{\ttfamily hep-th/9403137}}].

\bibitem{Calabrese:2009ez}
P.~Calabrese, J.~Cardy and E.~Tonni, \emph{{Entanglement entropy of two
  disjoint intervals in conformal field theory}},
  \href{https://doi.org/10.1088/1742-5468/2009/11/P11001}{\emph{J. Stat. Mech.}
  {\bfseries 0911} (2009) P11001}
  [\href{https://arxiv.org/abs/0905.2069}{{\ttfamily 0905.2069}}].

\bibitem{Hayden:2007cs}
P.~Hayden and J.~Preskill, \emph{{Black holes as mirrors: Quantum information
  in random subsystems}},
  \href{https://doi.org/10.1088/1126-6708/2007/09/120}{\emph{JHEP} {\bfseries
  09} (2007) 120} [\href{https://arxiv.org/abs/0708.4025}{{\ttfamily
  0708.4025}}].

\bibitem{Dieks:1982dj}
D.~Dieks, \emph{{COMMUNICATION BY EPR DEVICES}},
  \href{https://doi.org/10.1016/0375-9601(82)90084-6}{\emph{Phys. Lett. A}
  {\bfseries 92} (1982) 271}.

\bibitem{Russo:1992ax}
J.G.~Russo, L.~Susskind and L.~Thorlacius, \emph{{The Endpoint of Hawking
  radiation}}, \href{https://doi.org/10.1103/PhysRevD.46.3444}{\emph{Phys. Rev.
  D} {\bfseries 46} (1992) 3444}
  [\href{https://arxiv.org/abs/hep-th/9206070}{{\ttfamily hep-th/9206070}}].

\bibitem{Almheiri:2019yqk}
A.~Almheiri, R.~Mahajan and J.~Maldacena, \emph{{Islands outside the horizon}},
   \href{https://arxiv.org/abs/1910.11077}{{\ttfamily 1910.11077}}.

\bibitem{Anninos:2018svg}
D.~Anninos, D.A.~Galante and D.M.~Hofman, \emph{{De Sitter horizons \&
  holographic liquids}},
  \href{https://doi.org/10.1007/JHEP07(2019)038}{\emph{JHEP} {\bfseries 07}
  (2019) 038} [\href{https://arxiv.org/abs/1811.08153}{{\ttfamily
  1811.08153}}].

\bibitem{Aalsma:2020aib}
L.~Aalsma and G.~Shiu, \emph{{Chaos and complementarity in de Sitter space}},
  \href{https://doi.org/10.1007/JHEP05(2020)152}{\emph{JHEP} {\bfseries 05}
  (2020) 152} [\href{https://arxiv.org/abs/2002.01326}{{\ttfamily
  2002.01326}}].

\bibitem{Blommaert:2020tht}
A.~Blommaert, \emph{{Searching for butterflies in dS JT gravity}},
  \href{https://arxiv.org/abs/2010.14539}{{\ttfamily 2010.14539}}.

\bibitem{Chapman:2021eyy}
S.~Chapman, D.A.~Galante and E.D.~Kramer, \emph{{Holographic complexity and de
  Sitter space}}, \href{https://doi.org/10.1007/JHEP02(2022)198}{\emph{JHEP}
  {\bfseries 02} (2022) 198}
  [\href{https://arxiv.org/abs/2110.05522}{{\ttfamily 2110.05522}}].

\bibitem{Jorstad:2022mls}
E.~J\o{}rstad, R.C.~Myers and S.-M.~Ruan, \emph{{Holographic complexity in
  dS$_{d+1}$}}, \href{https://doi.org/10.1007/JHEP05(2022)119}{\emph{JHEP}
  {\bfseries 05} (2022) 119}
  [\href{https://arxiv.org/abs/2202.10684}{{\ttfamily 2202.10684}}].

\end{thebibliography}\endgroup

\end{document}